\documentclass[12pt]{amsart}

\usepackage{a4wide,amsmath,amssymb,graphicx}
\usepackage{color}

\let\pa\partial
\let\na\nabla
\let\eps\varepsilon
\newcommand{\R}{\mathbb{R}}

\newcommand{\N}{\mathbb{N}}

\newcommand{\diver}{\textnormal{div}}

\usepackage{graphicx}
\usepackage{caption}
\usepackage[labelformat=simple,labelfont={}]{subcaption}

\begin{document}



%
%
%

\title[Transient Schr\"odinger-Poisson simulations]
{
Transient Schr\"odinger-Poisson Simulations of a High-Frequency 
Resonant Tunneling Diode Oscillator
}
\author{Jan-Frederik Mennemann}
\address{Institute for Analysis and Scientific Computing, Vienna University of  
Technology, Wiedner Hauptstra\ss e 8--10, 1040 Wien, Austria}
\email{mennemann@asc.tuwien.ac.at}

\author{Ansgar J\"ungel}
\address{Institute for Analysis and Scientific Computing, Vienna University of  
Technology, Wiedner Hauptstra\ss e 8--10, 1040 Wien, Austria}
\email{juengel@tuwien.ac.at} 

\author{Hans Kosina}
\address{Institute for Microelectronics, Vienna niversity of Technology, 
Gusshausstra{\ss}e 27--29, 1040 Wien, Austria}
\email{kosina@iue.tuwien.ac.at}

\date{\today}

\thanks{
The first two authors acknowledge partial support from   
the Austrian Science Fund (FWF), grants P20214, P22108, and I395, and    
the Austrian-French Project of the Austrian Exchange Service (\"OAD).
}

\begin{abstract}
Transient simulations of a resonant tunneling diode oscillator are presented.
The semiconductor model for the diode consists of a set of time-dependent
Schr\"odinger equations coupled to the Poisson equation for the electric potential.
The one-dimensional Schr\"odinger equations are discretized by the finite-difference
Crank-Nicolson scheme using memory-type transparent boundary conditions which model
the injection of electrons from the reservoirs.
This scheme is unconditionally stable and reflection-free at the boundary.
An efficient recursive algorithm due to Arnold, Ehrhardt, and Sofronov is
used to implement the transparent boundary conditions,
enabling simulations which involve a very large number of time steps.
Special care has been taken to provide a discretization of the boundary data
which is completely compatible with the underlying finite-difference scheme.
The transient regime between two stationary states and the
self-oscillatory behavior of an oscillator circuit,
containing a resonant tunneling diode, is simulated for the first time.
\end{abstract}

%



\maketitle



\section{Introduction}

The resonant tunneling diode has a wide variety of applications as a
high-frequency and low-consumption oscillator or switch. 
The resonant tunneling structure is usually treated as an open quantum 
system with two large reservoirs and an active region containing 
a double-barrier heterostructure.
Accurate time-dependent simulations are of great importance to develop efficient 
and reliable quantum devices and to reduce their development time and cost. 
There exist several approaches in the literature to model a resonant tunneling diode. 
The simplest approach is to replace the diode by an equivalent circuit
containing nonlinear current-voltage characteristics \cite{KBHS11}.
Another approximation is to employ the Wannier envelope function development \cite{LLY96}. 
Other physics-based approaches rely on the Wigner equation \cite{Bie98,KKFR89},
the nonequilibrium Green's function theory \cite{Dat90,KLBFM95}, 
quantum hydrodynamic models \cite{Gar94,HTL08,JMM06},
and the Schr\"odinger equation \cite{BePi06,BFN09,Pin02,TaPo96}. 

In this paper, we adopt the latter approach and simulate the time-dependent behavior 
of a resonant tunneling diode using the Schr\"odinger-Poisson system in one space dimension. 
In this setting, the electrons are assumed to be in a mixed state with Fermi-Dirac statistics 
and the electrostatic interaction is taken into account at the Hartree level.
Each state is determined as the solution of the transient Schr\"odinger equation with 
nonhomogeneous transparent boundary conditions. 
The Schr\"odinger equations are discretized by the Crank-Nicolson finite-difference scheme and coupled self-consistently to the Poisson equation.
The main originality of this paper is the adaption (and slight improvement)
of existing numerical techniques from, e.g., \cite{AABES08,AES03,Pin02},
to a long-time simulation of a high-frequency oscillator circuit containing
a resonant tunneling diode. Our changes in the techniques are necessary
to achieve simulations without spurious oscillations 
in the numerical transient solution. In the following, we detail the techniques used
as well as the novel features.

{\em First}, we consider the one-dimensional
stationary problem, since it builds the
basis for the transient simulations. The stationary transparent boundary conditions
are discretized in such a way that their discrete version is compatible with
the underlying finite difference discretization, as proposed in \cite{Arn01}.
Thereby, any (numerical) spurious oscillation is eliminated, 
which would otherwise propagate in the transient simulations.
In the literature \cite{BeFa10,Pin02}, a modified version of the potential energy
is used to overcome problems of numerical convergence. Physically interpreted,
this model introduces artificial surface charge densities at the junction 
interfaces of the tunneling diode. We are able to solve the original problem.
This represents an improvement compared to the simulations in \cite{Pin02}, 
where the modified model is also employed for the time-dependent case.

{\em Second}, the time-dependent Schr\"odinger-Poisson system with
transparent boundary conditions is approximated. 
Since these boundary conditions are of memory type \cite{Arn01,BaPo91},
their numerical implementation requires to store (and to use) the boundary data
for all the past history. For this reason, simulations involving longer time
scales are extremely costly. This explains why simulations in the literature
\cite{BeFa10,BFN09,Pin02} have been restricted to some picoseconds only.
We solve this problem by using a fast evaluation of the discrete convolution
kernel of sum-of-exponentials, which has been presented in \cite{AES03}
and employed in \cite{AESS12} on circular domains.
To our knowledge, this rather new numerical technique has not been applied to
realistic device simulations so far.

A challenge in the transient simulations results from the large number
of wave functions which need to be propagated, accounting for the energy
distribution of the incoming electrons. Each state is provided with transparent
boundary conditions, which raises the computational cost sharply. To cope with
a large number of Schr\"odinger equations to be solved, we developed a parallel
version of our solver utilizing multiple cores on shared memory processors.
This enables us to present, for the first time, 
simulations to the Schr\"odinger-Poisson system
for large times up to 100\,ps (ps = picosecond) with reasonable computational
effort (compared to 5\,ps in \cite{Pin02}, 6\,ps in \cite{BFN09}, and 8\,ps
in \cite{BeFa10}).

Another novelty in this paper concerns the
discretization of nonhomogeneous discrete transparent boundary conditions.
They are necessary to describe continuously varying applied potentials
(as in an oscillator circuit). It is well known that, using a suitable gauge 
change, one can get rid of the transient potential \cite{AnBe03}.
Corresponding nonhomogeneous transparent boundary conditions can be found in
\cite{BeFa10}. In numerical simulations, however, we observed that these
boundary conditions may lead to unphysical distortions in the conduction
current density. The reason is that the considered discretization of the gauge change
is not compatible with the underlying finite-difference scheme. Therefore,
we suggest a new discretization which is derived from the Crank-Nicolson time
integration scheme. Our approach completely removes these numerical artifacts,
and we show that the total current density is now perfectly conserved.
We stress the fact that our discretization is completely consistent 
with the underlying Crank-Nicolson scheme inheriting its conservation and
stability properties.

{\em Third}, the numerical results allow us to identify plasma oscillations
in a certain time regime of the resonant tunneling diode and to estimate the life 
time of the resonant state. We present, for the first time, simulations of a
high-frequency oscillator circuit containing a reconant tunnelding diode, based
on a full Schr\"odinger-Poisson solver with transparent boundary conditions.
Simplified tunneling diode oscillators have been considered in 
\cite{KBHS11,LLY96,MuOkWa}. Our approach enables us to observe the
complex spatio-temporal behavior of macroscopic quantities inside the
resonant tunneling diode in an unprecedented way.

The paper is organized as follows. 
In Section \ref{sec.stat}, we detail the algorithm of the stationary problem.
The transient algorithm is described in Section \ref{sec.trans}.
In Section \ref{sec.numer} we consider numerical experiments for constant applied voltage and time-dependent applied voltage. 
Furthermore,
the numerical convergence related to the approximation of the discrete 
convolution kernel by sum-of-exponentials is investigated.
Finally, we present high-frequency oscillator circuit simulations in Section \ref{sec.circuit}.


\section{Stationary simulations}\label{sec.stat}

The steady state is the basis for the transient simulations. Therefore,
we discuss first the stationary regime. 

\subsection{Schr\"odinger-Poisson model}\label{sec.stat.model}

\begin{figure}[ht]
\begin{centering}
\includegraphics[width=150mm]{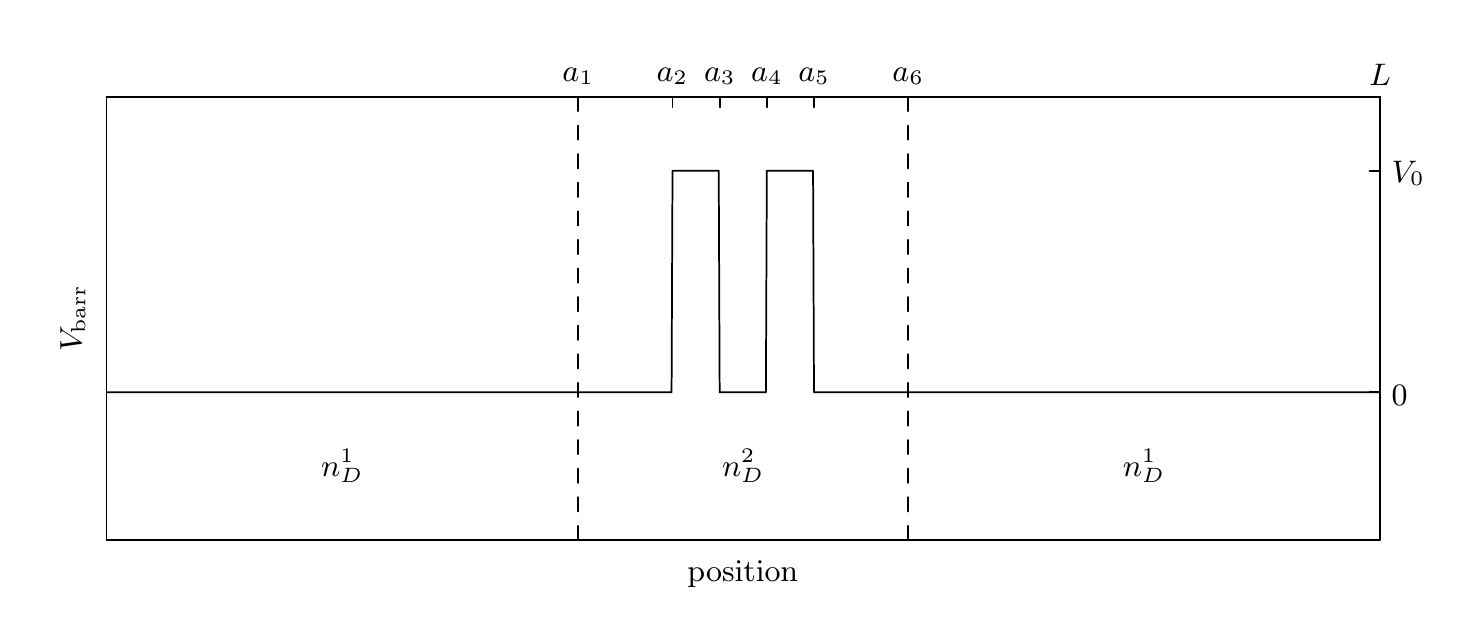}
\caption{Barrier potential and doping profile of a double-barrier heterostructure.}
\label{fig.rtd_layout}
\end{centering}
\end{figure}

We assume that the one-dimensional device in $(0,L)$ is connected
to the semi-infinite leads $(-\infty,0]$ and $[L,\infty)$. 
The leads are assumed to be in thermal
equilibrium and at constant potential. At the contacts, electrons are
injected with some given profile. We suppose that the charge transport is 
ballistic and that the electron wave functions evolve independently from each other. 
The one-dimensional device consists of three regions: two highly doped regions,
$[0,a_1]$ and $[a_6,L]$, with the doping concentration $n_D^1$
and a lowly doped region, $[a_1,a_6]$, with the doping density $n_D^2$ (see Figure
\ref{fig.rtd_layout}). The middle interval contains a double barrier,
described by the barrier potential
$$
  V_{\mathrm{barr}}(x) =
  \begin{cases}
  \begin{aligned}
  V_0 \quad &\text{for } x \in [a_2, a_3] \cup [a_4, a_5], \\
  0   \quad &\text{else}.
  \end{aligned}
  \end{cases}
$$
The doping profile $n_D$ is defined by 
$$
  n_D(x) = 
  \begin{cases}
  \begin{aligned}
  n_D^1 \quad &\text{for } x \in [0,a_1] \cup [a_6,L], \\
  n_D^2 \quad &\text{else}.
  \end{aligned}
  \end{cases}
$$
The parameters are taken from \cite{BeFa10,Pin02}:
\begin{center}
  \begin{tabular}{llll}
  $a_1=50$\,nm, & $a_2=60$\,nm, & $a_3=65$\,nm, \\
  $a_4=70$\,nm, & $a_5=75$\,nm, & $a_6=85$\,nm, \\
  $L=135$\,nm, & $n_D^1=10^{24}$\,m$^{-3}$, & $n_D^2=5\cdot 10^{21}$\,m$^{-3}$,
  \end{tabular}
\end{center}
and the barrier height is $V_0=0.3$\,eV. 

The Coulomb interaction is taken into account at the Hartree level, i.e.\ by
an infinite number of Schr\"odinger equations
\begin{equation}\label{eq.stationary_schroedinger_equation}
  -\frac{\hbar^2}{2m}\,\frac{d^2\phi_k}{dx^2}(x) + V(x) \phi_k(x) 
  = E(k) \phi_k(x), \quad x \in \R,
\end{equation}
self-consistently coupled to the Poisson equation,
\begin{equation}\label{eq.V_self}
\begin{aligned}
  -\frac{d^2 V_{\rm self}}{dx^2} 
  &= \frac{e^2}{\eps} (n[V_{\mathrm{self}}] - n_D),\quad x \in (0,L), \\ 
  V_{\mathrm{self}}(0) &= 0, \quad V_{\mathrm{self}}(L) = - e U,
\end{aligned}
\end{equation}
where $V=V_{\rm barr}+V_{\rm self}$ is the potential energy.
The physical parameters are the reduced Planck constant $\hbar$,
the effective electron mass $m$, the elementary charge $e$, and the
permittivity $\eps=\eps_r\eps_0$, being the product of the relative permittivity
$\eps_r$ and the electric constant $\eps_0$. Furthermore, $U\ge 0$ denotes
the applied voltage, and the electron density is defined by
\begin{equation}\label{eq.electron_density}
  n[V_{\mathrm{self}}](x) = \int_{\R} g(k) |\phi_k(x)|^2 dk.
\end{equation}
The injection profile $g(k)$ is given according to Fermi-Dirac statistics by 
\begin{equation}\label{fermi.dirac}
  g(k) = \frac{m k_B T_0}{2 \pi^2 \hbar^2}\ln 
  \left(1 + \exp\left(\frac{E_F - \hbar^2 k^2/(2m)}{k_B T_0}\right)\right),
\end{equation}
where $k_B$ is the Boltzmann constant, $T_0$ the temperature of the
semiconductor and $E_F$ the Fermi energy (relative to the conduction band edge).
In all subsequent simulations, we use, as in \cite{Pin02},
$\eps_r = 11.44$, $T_0 = 300$\,K, $E_F = 6.7097\cdot 10^{-21}$\,J, 
and the effective mass of Gallium arsenide, $m = 0.067m_e$, with $m_e$ being
the electron mass at rest.

In order to define the total electron energy $E(k)$ depending on the wave number
$k\in\R$, we need to distinguish the cases $k>0$ and $k<0$. For $k>0$,
the electrons enter from the left, and we have
$E(k)=\hbar^2k^2/(2m)$. The wave function in the leads is given by
\begin{align*}
  & \phi_k(x) = e^{i k x} + r(k) e^{-i k x},\quad x < 0, \\
  & \phi_k(x) = t(k)\exp\left(i \sqrt{2 m (E(k) - V(L))/\hbar^2}x\right), \quad x > L.
\end{align*}
Eliminating the transmission and reflection coefficients $t(k)$ and $r(k)$, 
respectively, the boundary conditions
\begin{equation}\label{eq.analytic_stationary_boundary_conditions_1}
  \phi_k'(0) + i k \phi_k(0) = 2 i k, \quad
  \phi_k'(L) = i \sqrt{2m (E(k)-V(L)) / \hbar^2} \phi_k(L)
\end{equation}
follow. For $k<0$, the electrons enter from the right. The total energy is
given by $E(k)=\hbar^2k^2/(2m)-eU$, and the wave function in the leads reads as
\begin{align*}
  & \phi_k(x) = t(k) \exp\left(- i \sqrt{2 m E(k) / \hbar^2} x\right), \quad x < 0, \\
  & \phi_k(x) = e^{i k x} + r(k) e^{-i k x},\quad x > L.
\end{align*}
This yields the boundary conditions
\begin{equation}\label{eq.analytic_stationary_boundary_conditions_2}
  \phi_k'(0) = - i \sqrt{2 m E(k) / \hbar^2} \phi_k(0), \quad
  \phi_k'(L) + i k \phi_k(L) = 2 i k e^{i k L}. 
\end{equation}

Summarizing, the stationary problem consists in the
Schr\"odinger equation \eqref{eq.stationary_schroedinger_equation} 
with the transparent boundary conditions 
\eqref{eq.analytic_stationary_boundary_conditions_1}-%
\eqref{eq.analytic_stationary_boundary_conditions_2} coupled to the
Poisson equation \eqref{eq.V_self} via the electron density 
\eqref{eq.electron_density}. 
We remark that the existence and uniqueness of solutions to a 
Schr\"odinger-Poisson boundary-value problem similar to 
\eqref{eq.stationary_schroedinger_equation}-\eqref{eq.analytic_stationary_boundary_conditions_2}
has been shown in \cite{BeDeMa}.


\subsection{Discrete transparent boundary conditions}\label{sec.stat.tbc}

We recall the finite-difference discretization of the stationary Schr\"odinger equation
with transparent boundary conditions \cite{Arn01}. Using standard second-order
finite differences on the equidistant grid $x_j=j\triangle x$, 
$j\in\{0,\ldots,J\}$,
with $x_J=L$ and $\triangle x>0$, we find for the grid points located in the
computational domain,
\begin{equation}\label{eq_discretized_stationary_schroedinger}
  \phi_{j+1} - 2\phi_j + \phi_{j-1} 
  + \frac{2m(\triangle x)^2}{\hbar^2} (E(k)-V_j) \phi_j = 0.
\end{equation}

It is well known that a standard centered finite-difference discretization
of the boundary conditions \eqref{eq.analytic_stationary_boundary_conditions_1}
and \eqref{eq.analytic_stationary_boundary_conditions_2} may lead to spurious
oscillations in the numerical solution \cite{Arn01}. In principle, the numerical
errors can be made as small as desired by choosing $\triangle x$ sufficiently
small. However, since the stationary solutions will serve as intitial states in our 
transient simulations, we need to avoid any spurious oscillation, which would 
otherwise be propagated with every time step.

For this, we apply (stationary) discrete transparent boundary conditions
compatible with the finite-difference discretization \eqref{eq_discretized_stationary_schroedinger} as proposed in \cite{Arn01}.
For the sake of completeness, we review the derivation.
Note that the final discretization is equivalent to the discretization \eqref{eq_discretized_stationary_schroedinger}
extended to the whole space, i.e.\ for $j \in \mathbb{Z}$.

In the semi-infinite leads $j\le 0$ and $j\ge J$, 
the potential energy is assumed to be constant,
$$
  V_j =
  \begin{cases}
  \begin{aligned}
  V_0 &= 0  \quad &\text{for } j \leq 0, \\
  V_J &=-eU \quad &\text{for } j \geq J.
  \end{aligned}
  \end{cases}
$$
Then \eqref{eq_discretized_stationary_schroedinger} reduces to a difference equation with constant coefficients which admits two solutions of the form
$\phi_j= (\alpha_{0,J}^\pm)^j$, where
\[
\begin{aligned}
  \alpha_{0,J}^{\pm} &= 1 - \frac{m (E(k)-V_{0,J}) (\triangle x)^2}{\hbar^2} \\
  &\phantom{xx}{}\pm i \sqrt{\frac{2 m (E(k)-V_{0,J}) (\triangle x)^2}{\hbar^2}
  - \frac{m^2 (E(k)-V_{0,J})^2 (\triangle x)^4}{\hbar^4}}.
\end{aligned}
\]
Here, $E(k)-V_{0,J}$ corresponds to the kinetic energy $E^{\mathrm{kin}}_{0,J}(k)$ in the left or right lead.
In case $E^{\mathrm{kin}}_{0,J}(k) > 0$, the solution is a discrete plane wave and 
$(\triangle x)^2<2\hbar^2/(m ( E(k) - V_{0,J}))$ is needed to ensure 
$|\alpha_{0, J}|=1$, 
which in practise is not a restriction.
In case $E^{\mathrm{kin}}_{0,J}(k) = 0$, the solution is constant.
Depending on the applied voltage, $E^{\mathrm{kin}}_{0,J}(k)$ might also become negative.
In that case, the solution is decaying or growing exponentially fast and we select the decaying solution 
as it is the only physically reasonable solution.

In practice, we start with the calculation of the total energy $E(k) = E^{\mathrm{kin}}_{0,J}(k) + V_{0,J}$.
For electrons coming from the left contact we have $E(k) = E^{\mathrm{kin}}_{0}(k)$.
As the incoming electron is represented by a discrete plane wave, $E^{\mathrm{kin}}_{0}(k)$ is positive but, depending on the applied voltage,
$E^{\mathrm{kin}}_J(k)$ might be positive, zero or negative.
For electrons coming from the right contact, 
we have $E(k) = E^{\mathrm{kin}}_J(k) - e U$.
Again, the incoming wave function is a discrete plane wave, i.e., $E^{\mathrm{kin}}_J(k) > 0$ but nothing is said about
$E^{\mathrm{kin}}_0(k)$.
At this point it should be noted that the kinetic energy of the incoming electron needs to be computed according to the discrete dispersion relation
$$
  E^{\mathrm{\mathrm{kin}}}(k) = \frac{\hbar^2}{m (\triangle x)^2}( 1 - \cos(k \triangle x)),
$$
which follows after solving the centered finite-difference discretization of the free Schr\"o\-din\-ger equation
$$
  -\frac{\hbar^2}{2m}\,\frac{d^2}{dx^2}e^{ikx}=E^{\mathrm{kin}}(k) e^{ikx}.
$$  
In the limit $\triangle x\to 0$, we recover the continuous dispersion
relation $E^{\mathrm{kin}}(k)=\hbar^2k^2/(2m)$. 

Let us consider a wave function entering the device from the left contact ($k>0$).
For $j\le 0$, the solution to \eqref{eq_discretized_stationary_schroedinger}
is a superposition of an incoming and a reflected discrete plane wave,
$\phi_j=\beta^j+B\beta^{-j}$, where $\beta = \alpha_0$.
We eliminate $B$ from $\phi_{-1}=\beta^{-1}+B\beta$, $\phi_0=1+B$ to find 
the discrete transparent boundary condition at $x_0$:
\begin{equation*}
  -\beta^{-1} \phi_{-1} + \phi_0 = 1 - \beta^{-2}.
\end{equation*}
For $j\ge L$, the solution to \eqref{eq_discretized_stationary_schroedinger}
is given by $\phi_j=C\gamma^j$ with $\gamma = \alpha_J$.
This means that $\phi_{J+1}=C\gamma^{J+1}=\gamma\phi_J$, and the boundary condition
at $x_J$ becomes
$$
  \phi_J - \gamma^{-1} \phi_{J+1} = 0.
$$

Summarizing, we obtain the linear system $A\phi=b$ with the tridiagonal matrix
$A$ consisting of the main diagonal $(-\beta^{-1},-2+2m(\triangle x)^2(E(k)-V_0)/\hbar^2,
\ldots,-2+2m(\triangle x)^2(E(k)-V_J)/\hbar^2,-\gamma^{-1})$ and the first off
diagonals $(1,\ldots,1)$.
The vector of the unknowns is given by $\phi=(\phi_{-1},\ldots,\phi_{J+1})^\top$ and $b$ represents the right-hand side
$b=(1-\beta^{-2},0,\ldots,0)^\top$.

The case of a wave function entering from the right contact ($k<0$) works analogously.


\subsection{Solution of the Schr\"odinger-Poisson system}\label{sec.stat.sp}

We explain our strategy to solve the coupled Schr\"odinger-Poisson system.
To this end, we introduce the equidistant energy grid
\begin{equation}\label{2.K}
  {\mathcal K} = \{-k_M,-k_M+\triangle k,\ldots,-\triangle k, +\triangle k,\ldots, k_M-\triangle k,k_M\}, \quad
  K:=|{\mathcal K}|.
\end{equation}
The electron density \eqref{eq.electron_density} is approximated by
$$
  n_{\rm disc}[V_{\rm self}](x) = \triangle k\sum_{k\in{\mathcal K}}g(k)|\phi_k(x)|^2,
$$
where the Fermi-Dirac statistics $g(k)$ is defined in \eqref{fermi.dirac}
and the functions $\phi_k$ are the scattering states,
i.e., the solutions to the discretized stationary Schr\"odinger equation \eqref{eq_discretized_stationary_schroedinger}
with discrete transparent boundary conditions as described in
Section \ref{sec.stat.tbc}.
This approximation is reasonable if
$\triangle k$ is sufficiently small and $k_M$ is sufficiently large.
In the numerical simulations below, we choose $K=3000$ and, as in 
\cite[Section 5]{BeFa10}, $k_M=\sqrt{2m(E_F+7k_B T_0)}/\hbar$, recalling that
$E_F=6.7097\cdot 10^{-21}$\,J and $T_0=300$\,K. 

The discrete Schr\"odinger-Poisson system is iteratively solved as follows.
We set $V=V_{\rm barr}+V_{{\rm self},U}^{(p)}$, where $V_{{\rm self},U}^{(p)}$
is the $p$-th iteration of $V_{\rm self}$ for the applied voltage $U$.
Given $V$, we compute a set of quasi eigenstates 
$\{\phi_k^{(p)}\}_{k\in{\mathcal K}}$.
This defines the discrete electron density
$$
  n_{\rm disc}[V_{{\rm self},U}^{(p)}] = \triangle k\sum_{k\in{\mathcal K}}g(k)
  |\phi_k^{(p)}(x)|^2.
$$
The Poisson equation is solved by employing a Gummel-type method \cite{Gu1964}:
  \begin{align*}
  & -\frac{d^2}{dx^2} V_{\mathrm{self},U}^{(p+1)} 
  = \frac{e^2}{\eps}\left(n[V_{\mathrm{self},U}^{(p)}] 
  \exp\left(\frac{V_{\mathrm{self},U}^{(p)}
  -V_{\mathrm{self},U}^{(p+1)}}{V_{\mathrm{self}}^{\mathrm{ref}}}\right) 
  - n_D\right),
  \\
  & V_{\mathrm{self},U}^{(p+1)}(0) = 0, \quad
  V_{\mathrm{self},U}^{(p+1)}(L) = -e U. 
\end{align*}
The idea of the Gummel method is to decouple the Schr\"odinger and
Poisson equations but to formulate the Poisson equation in a nonlinear way,
using the relation between the electron density and electric potential in
thermal equilibrium.
The parameter $V_{\rm self}^{\rm ref}$ can be tuned to reduce the number
of iterations; we found empirically that the choice 
$V_{\rm self}^{\rm ref}=0.04$\,eV minimizes the iteration number.
If the relative error in the $\ell^2$-norm is smaller than a fixed tolerance,
\begin{equation}\label{eq.gummel_stop}
  \left\|\frac{V_{\mathrm{self},U}^{(p+1)}
  -V_{\mathrm{self},U}^{(p)}}{V_{\mathrm{self},U}^{(p+1)}}\right\|_2 
  \le \delta, 
\end{equation}  
we accept
$V_{\rm self}:=V_{\mathrm{self},U}^{(p+1)}$ and $\{\phi_k^{(p+1)}\}_{k\in
{\mathcal K}}$ as the approximate self-consistent solution. 
Otherwise, we proceed with the iteration $p+1 \to p+2$ and compute a new set
of scattering states. The procedure is repeated until \eqref{eq.gummel_stop}
is fulfilled. We have choosen the tolerance $\delta=10^{-6}$.

For zero applied voltage we use $V_{{\rm self},0\,{\rm mV}}^{(0)}=0$\,mV to start the iteration. 
Only 7 iterations are needed until criterion \eqref{eq.gummel_stop} is fulfilled.
As a result we obtain $V_{{\rm self},0\,{\rm mV}}^{(7)}$, which is depicted in
Figure \ref{fig:V_self_divergence_and_V_self_Pinaud} (left part, solid line). 

Numerical problems arise when non-equilibrium solutions are computed.
As an example we consider the case of a small applied voltage $U=1$\,mV.
To start the iteration process we use the previously computed solution, i.e., we 
set $V_{{\rm self},1\,\mathrm{mV}}^{(0)}:=V_{{\rm self},0\,\mathrm{mV}}^{(7)}$.
The next iterations are illustrated in
Figure \ref{fig:V_self_divergence_and_V_self_Pinaud} (left part, dashed lines). 
Obviously they do not converge and are physically not realistic.
This phenomenon is a well-known in the literature \cite{Fis98,Fre89} and
is believed to be related to the absence of inelastic processes in
the Schr\"odinger-Poisson equations.

In the literature \cite{BeFa10,Pin02}, a modified version of the 
Schr\"odinger-Poisson equations is employed to overcome this problem.
The modification concerns the description of the potential energy in the Poisson equation.
For this, we write the Poisson equation \eqref{eq.V_self} as follows:
\begin{align*}
  & -\frac{d^2 V_0}{dx^2} = 0\phantom{\frac{e^2}{\eps}(n-n_D)} 
  \quad\mbox{in }(0,L), \quad
  V_0(0)=0, \quad V_0(L)=-eU, \\
  & -\frac{d^2 V_1}{dx^2} = \phantom{0}\frac{e^2}{\eps}(n-n_D)
  \quad\mbox{in }(0,L), \quad
  V_1(0)=0, \quad V_1(L)=0,
\end{align*}
i.e., the self-consistent potential is $V_{\rm self}=V_0+V_1$.
The first boundary-value problem can be solved explicitly:
$V_0(x)=-eUx/L$, $x\in[0,L]$. In \cite{BeFa10,Pin02}, the linearly
decreasing potential $V_0$ has been replaced by the ramp-like potential
\begin{equation}\label{3.V1.iter}
  \widetilde V_0(x) = -B_0\left(\frac{x-a_1}{a_6-a_1}\mathbf{1}_{[a_1,a_6)}
  + \mathbf{1}_{[a_6,\infty)}\right), \quad x\in[0,L],
\end{equation}
where $\mathbf{1}_I$ is the characteristic function on the interval $I\subset\R$ 
(see Figure \ref{fig.rtd_layout} for the definition of $a_1$ and $a_6$).
The function $\widetilde V_0+V_{\rm barr}$ is illustrated in Figure 
\ref{fig:V_self_divergence_and_V_self_Pinaud} (right part, dotted line).
The potential energy is then given by $V=\widetilde V_0+V_1+V_{\rm barr}$.
Using this modified physical model,
the above Gummel iteration scheme for the Poisson equation
for $V_1$ converges without any problems, see Figure \ref{fig:V_self_divergence_and_V_self_Pinaud} 
(right part, solid line), even for large applied voltages.
However, we will see below that the results from the modified model differ considerably 
from the results obtained by the original Schr\"odinger-Poisson model.
Furthermore, the potential energy is no longer differentiable at $a_1$ and $a_6$. 
This may be interpreted as a model of surface charge densities at the
interfaces which, however, are not intended in the model.

\begin{figure}
\label{fig:V_self_divergence_and_V_self_Pinaud}
\includegraphics[width=75mm]{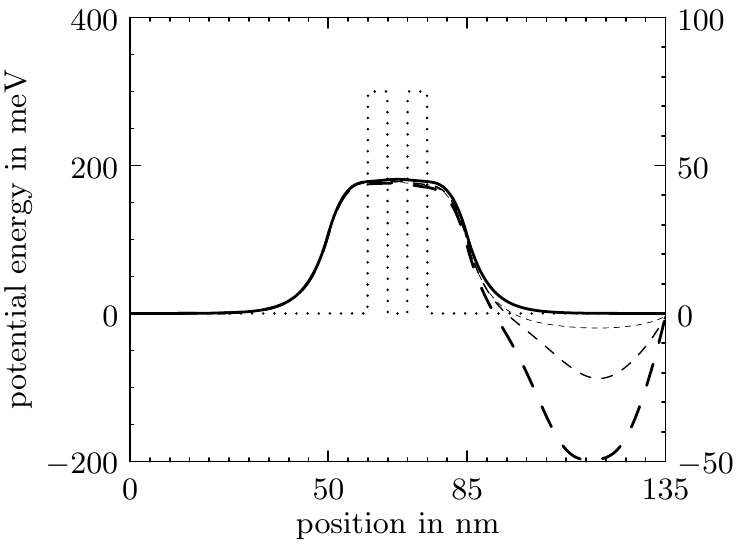}
\hfill
\includegraphics[width=75mm]{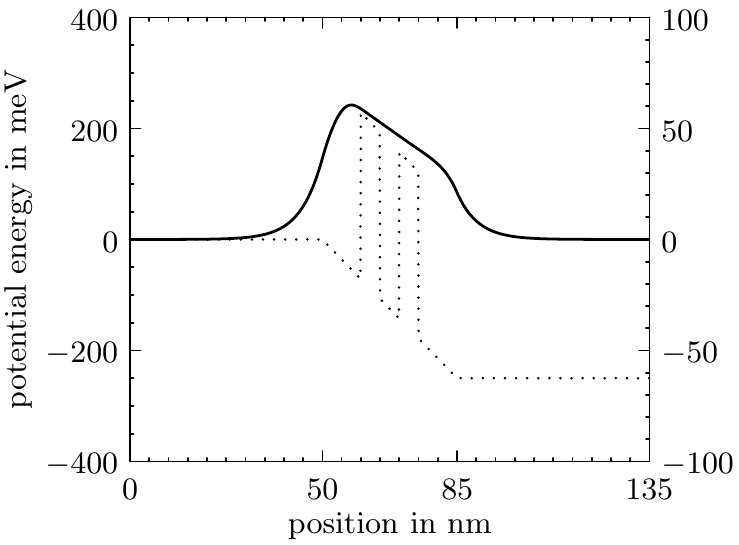}
\caption{
Left part: Solid line: self-consistent solution $V_{{\rm self}}$ for $U=0$\,mV, found after
7 iterations. Dashed lines: divergent approximations for $U=1$\,mV.
Dotted line: barrier potential.
Right part: Solid line: self-consistent solution $V_1$ for $U=250$\,mV according to approximation \eqref{3.V1.iter}.
Dotted line: sum of the barrier potential and a ramp-like potential.}
\label{fig:V_self_divergence_and_V_self_Pinaud}
\end{figure}

\begin{figure}[htb]
\begin{centering}
\includegraphics[width=140mm]{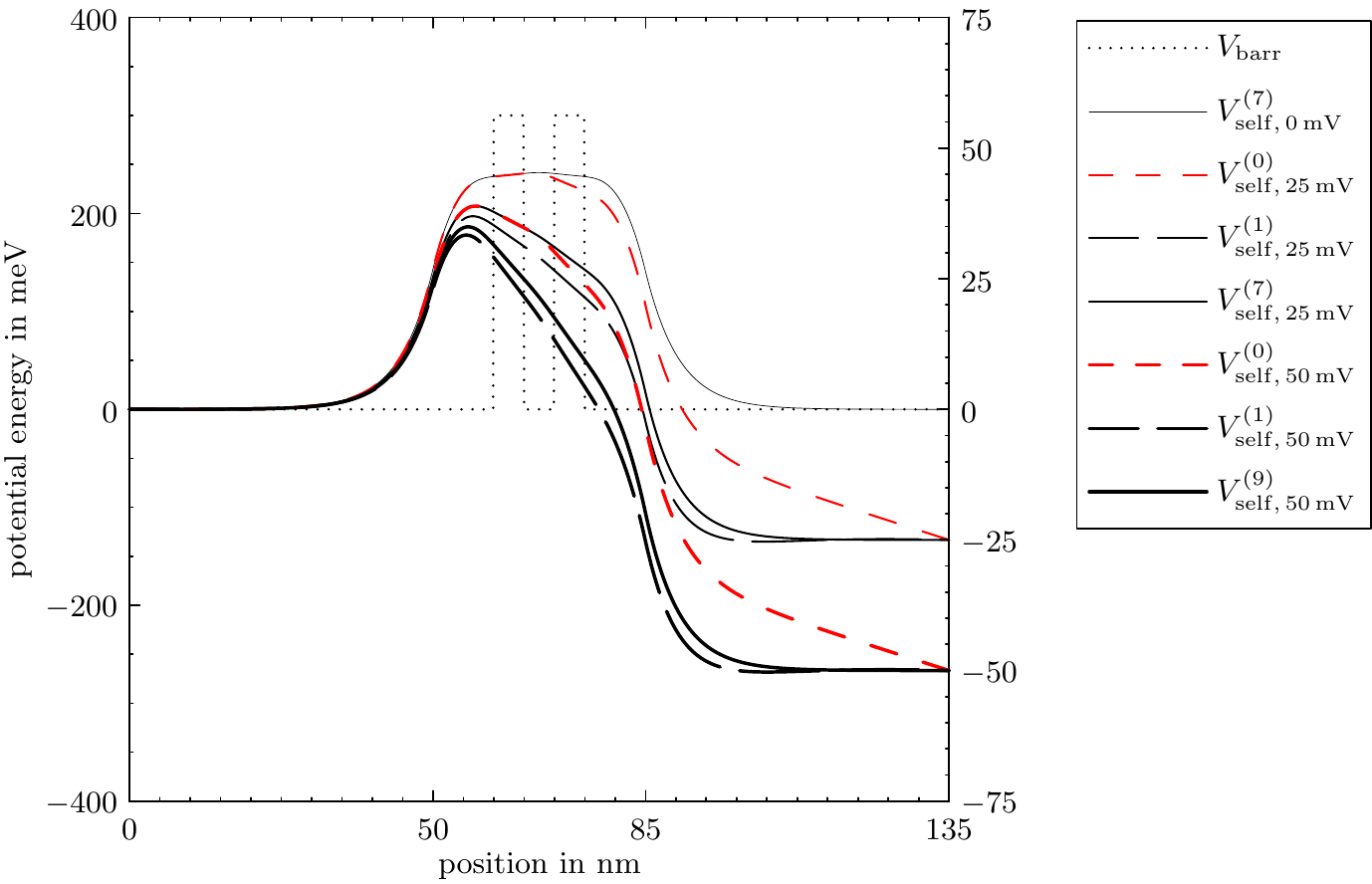}
\caption{Some iterations computed according to \eqref{3.Vself.iter}.}
\label{fig.V_self_convergence}
\end{centering}
\end{figure}

In fact, we are able to solve numerically 
the original Schr\"odinger-Poisson problem.
To this end, the applied voltage needs to be increased in small steps.
We found that the starting potential in each step needs to be initialized carefully. 
More precisely, given the self-consistent solution $V_{\mathrm{self,U}}$
for the applied voltage $U$, we wish to compute a self-consistent
solution with the applied voltage $U+\triangle U$.
In each step we choose
\begin{equation}
\label{3.Vself.iter}
  V_{{\rm self},U+\triangle U}^{(0)}(x):= V_{{\rm self},U}(x) 
  - \triangle U\frac{2x-L}{L}\mathbf{1}_{[L/2,L]}
\end{equation}
to start the iteration. For $U=0$\,mV and $\triangle U=25$\,mV,
the Gummel scheme converges to a physically reasonable 
solution after 7 iterations (i.e., \eqref{eq.gummel_stop} is fulfilled). 
Some iterations are shown in Figure \ref{fig.V_self_convergence}.
We observed that a voltage step $\triangle U<30$\,mV leads to convergent
solutions also for large applied voltages.

\begin{figure}[htb]
\begin{centering}
\includegraphics[width=150mm]{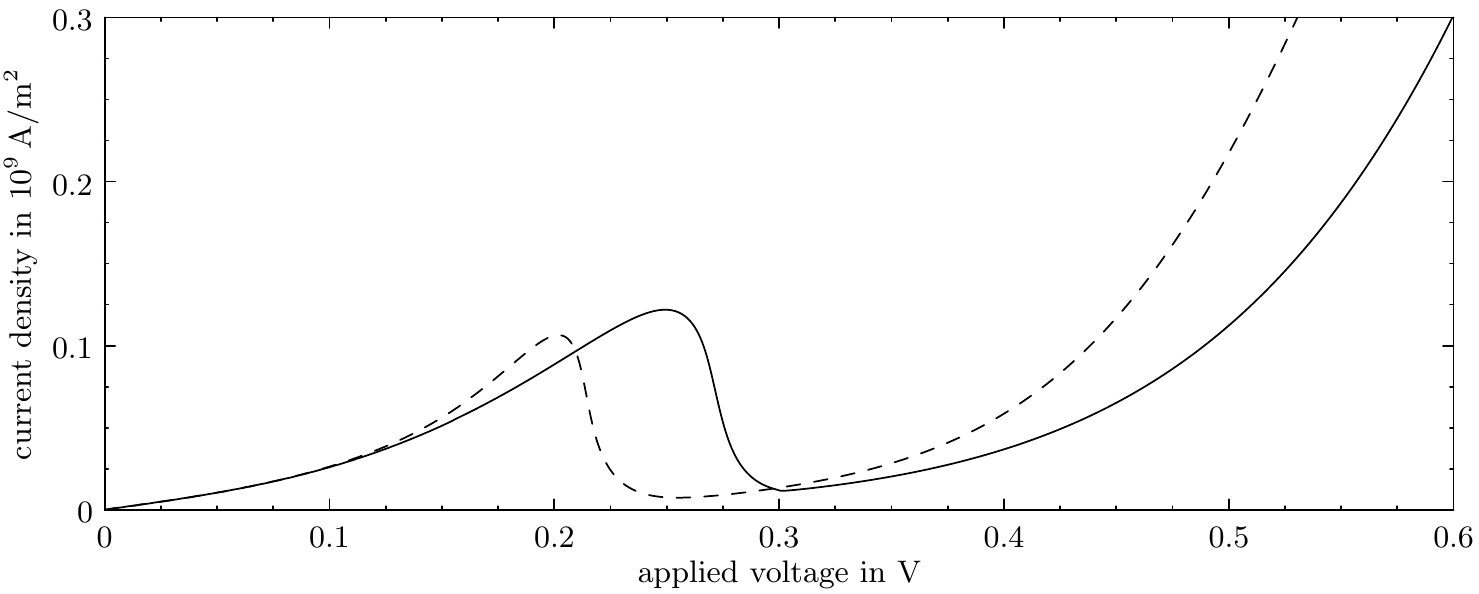}
\caption{Current-voltage characteristics. 
The solid line corresponds to 
our solution of the original stationary Schr\"odinger-Poisson system.
The dashed line is obtained with the
modified model using approximation \eqref{3.V1.iter}.}
\label{fig.current_density_of_voltage_curves}
\end{centering}
\end{figure}

In order to compare the original Schr\"odinger-Poisson model with the model 
using approximation \eqref{3.V1.iter},
we computed the current-voltage characteristics shown in
Figure \ref{fig.current_density_of_voltage_curves}.
Here, the (conduction) current density
\begin{equation}\label{2.J}
  J_{\rm cond} 
  = \frac{e\hbar}{m}\int_\R g(k)\mbox{Im}\left(\phi_k^* \frac{d\phi_k}{dx}\right)dk
\end{equation}
is approximated by a simple quadrature formula using
symmetric finite differences to compute $d\phi_k/dx$.
Figure \ref{fig.current_density_of_voltage_curves} shows that the results differ
considerably, i.e., the choice \eqref{3.V1.iter} leads to different results than
those computed from the original model. Therefore, 
we employ the original potential energy
in the transient simulations in the next subsection.


\section{Transient simulations}\label{sec.trans}

In this section, we detail the numerical discretization of the transient
Schr\"odinger equations 
\begin{equation}\label{3.se}
  i\hbar\frac{\pa\psi_k}{\pa t} = -\frac{\hbar^2}{2m}\,\frac{\pa^2\psi_k}{\pa x^2}
  + V(\cdot,t)\psi_k, \quad \psi_k(\cdot,0) = \phi_k, \quad
  x\in[0,L],\ t>0,\ k\in{\mathcal K}, 
\end{equation}
with discrete transparent boundary conditions, where
${\mathcal K}$ is defined in \eqref{2.K}. 
To simplify the presentation, we skip in this section the index $k$.


\subsection{Nonhomogeneous discrete transparent boundary conditions}
\label{sec.trans.dtbc}

The transient Schr\"odinger equation \eqref{3.se} is discretized by the commonly used
Crank-Nicolson scheme:
\begin{equation}\label{3.cn}
\begin{aligned}
  \psi_{j-1}^{(n+1)} + \Big( i R - 2 &+ w V_j^{(n+1/2)} \Big) \psi_j^{(n+1)} 
  + \psi_{j+1}^{(n+1)} \\
  &= - \psi_{j-1}^{(n)} + \Big( i R + 2 - w V_j^{(n+1/2)} \Big) \psi_j^{(n)} 
  - \psi_{j+1}^{(n)},
\end{aligned}
\end{equation}
where $\psi_j^{(n)}$ approximates $\psi(x_j,t_n)$ with $x_j=j\triangle x$ and
$t_n=n\triangle t$ ($j\in{\mathbb Z}$, $n\in\N_0$), 
$V_j^{(n+1/2)}$ approximates $V(j\triangle x,(n+1/2)\triangle t)$, 
and $R=4m(\triangle x)^2/(\hbar\triangle t)$, $w=-2m(\triangle x)^2/\hbar^2$.
Under the assumptions that the initial wave function is compactly supported
in $(0,L)$ and that the applied voltage vanishes, $V(x,t)=0$ for $x\le 0$ and $x\ge L$,
$t\ge 0$, it is well known (see, e.g., \cite{Arn01,BaPo91})
that transparent boundary conditions for the 
Schr\"odinger equation \eqref{3.se} read as
\begin{subequations}\label{eq.homogeneous_transparent_boundary_conditions}
  \begin{align}
  \frac{\pa \psi}{\pa x}(0,t) &= \sqrt{\frac{2 m}{\pi \hbar}} e^{-i \pi/4} 
  \frac{d}{dt}\int_0^t \frac{\psi(0,\tau)}{\sqrt{t-\tau}} \,d \tau,  
  \label{eq.homogeneous_transparent_boundary_conditions_1} \\ 
  \frac{\pa \psi}{\pa x}(L,t) &= -\sqrt{\frac{2 m}{\pi \hbar}} e^{-i \pi/4} 
  \frac{d}{dt} \int_0^t \frac{\psi(L,\tau)}{\sqrt{t-\tau}} \,d \tau. \label{eq.homogeneous_transparent_boundary_conditions_2}
\end{align}
\end{subequations}
The (homogeneous) discrete transparent boundary conditions, based on the above 
Crank-Nicolson scheme, are given as follows 
(see \cite{Arn98} for the derivation):
\begin{subequations}\label{eq.discrete_homogeneous_boundary_conditions}
  \begin{align}
  \psi_1^{(n+1)}-s^{(0)} \psi_0^{(n+1)} 
  &= \sum_{\ell=1}^{n} s^{(n+1-\ell)} \psi_0^{(\ell)} - \psi_1^{(n)}, \quad n\geq 0, 
  \label{eq.discrete_homogeneous_transparent_boundary_conditions_1}\\ 
  \psi_{J-1}^{(n+1)} - s^{(0)} \psi_J^{(n+1)} 
  &= \sum_{\ell=1}^{n} s^{(n+1-\ell)} \psi_J^{(\ell)} - \psi_{J-1}^{(n)}, \quad n \geq 0, 
  \label{eq.discrete_homogeneous_transparent_boundary_conditions_2}
\end{align}
\end{subequations}
with the convolution coefficients
\begin{equation}\label{eq.discrete_convolution_coefficients}
  s^{(n)} = \left( 1 - i \frac{R}{2} \right) \delta_{n,0} 
  + \left( 1 + i \frac{R}{2} \right) \delta_{n,1}
  + \alpha e^{-i n \varphi} \frac{P_n(\mu)-P_{n-2}(\mu)}{2n-1}
\end{equation}
and the abbreviations
$$
  \varphi = \arctan{ \frac{4}{R} },\quad 
  \mu = \frac{R}{\sqrt{R^2+16}}, \quad 
  \alpha = \frac{i}{2} \sqrt[4]{R^2(R^2+16)} e^{i \varphi / 2}.
$$
Here, $P_n$ denotes the $n$th-degree Legendre polynomial ($P_{-1}=P_{-2}=0$),
and $\delta_{n,j}$ is the Kronecker symbol.
In practice, the coefficients defined in
\eqref{eq.discrete_convolution_coefficients} 
are computed with an efficient three-term recursion, relying on the three-term 
recursion of the Legendre polynomials \cite{EhAr01}.
The Crank-Nicolson scheme along with
these discrete boundary conditions yields an unconditionally
stable discretization which is perfectly reflection-free \cite{Arn98,Arn01}.

Next, let the initial wave function be a solution to the stationary Schr\"odinger
equation with energy $E$ and let the exterior potential at the right contact
be given by a time-dependent function, $V(x,t)=-eU(t)$ for $x\ge L$, $t\ge 0$.
This leads to nonhomogeneous transparent boundary conditions \cite{AABES08}. 
We describe our strategy to discretize these boundary conditions.
Our approach is motivated by that presented in \cite[Appendix B]{BeFa10},
but we suggest, similarly as in \cite{Arn01}, a discretization of the
gauge change which is compatible with the underlying finite-difference scheme.
Additionally, our approach requires only a single set of convolution coefficients instead of two.

First, we derive a nonhomogeneous discrete transparent boundary condition
at $x_j=L$. To this end, we consider the difference between the unknown wave
function $\psi$ and the time evolution of the scattering state, $\phi(x)\exp(iEt/\hbar)$,
in the right lead $[L,\infty)$. 
We employ a gauge change to get rid of the time-dependent potential $V_L(t)=-eU(t)$. 
As a consequence, the function
$\psi(x,t)\exp(i\int_0^t V_L(s)ds/\hbar)$ solves the free transient Schr\"odinger
equation in $[L,\infty)$. Using a similar gauge change, a straightforward
computation shows that $\phi(x)\exp(-i(E-V_L(0))t/\hbar)$ solves the free
Schr\"odinger equation in $[L,\infty)$ as well. Hence,
\begin{equation}\label{3.phi}
  \varphi(x,t) := \psi(x,t)\exp\left(\frac{i}{\hbar}\int_0^t V_L(s)ds\right)
  - \phi(x)\exp\left(-\frac{i}{\hbar}(E-V_L(0))t\right)
\end{equation}
for $x\in[L,\infty)$
solves the free transient Schr\"odinger equation. 
Furthermore, $\varphi(x,0)=0$ for all $x\in[L,\infty)$. 
Therefore, we could apply \eqref{eq.homogeneous_transparent_boundary_conditions_2} 
to derive a nonhomogeneous transparent boundary condition at the right contact.
Instead we replace $\varphi(x,t)$ by some approximation $\varphi_j^{(n)}$ and subsequently 
apply \eqref{eq.discrete_homogeneous_transparent_boundary_conditions_2} 
to derive a discrete boundary condition compatible with the Crank-Nicolson scheme. 
The question is how to approximate the quantities
$\exp(i\int_0^t V_L(s)ds/\hbar)$ and $\exp(-i(E-V_L(0))t/\hbar)$. Indeed,
the ad-hoc discretization for $t=n \triangle t$,
\begin{equation}\label{eq.discrete_gauge_change_non_consistent}
\begin{aligned}
  \exp\left(\frac{i}{\hbar}\int_0^t V_L(s)ds\right)
  &\approx \exp\left(\frac{i}{\hbar}\sum_{\ell=0}^{n-1} V_L^{(\ell + 1/2)} 
  \triangle t\right),\\
  \exp\left(-\frac{i}{\hbar}(E-V_L(0))t\right)
  &= \exp\left(-\frac{i}{\hbar}(E-V_L^{(0)})n\triangle t\right),
\end{aligned}
\end{equation}
where $V_L^{(\ell)}=V_L(\ell \triangle t)$,
is not derived from the underlying finite-difference discretization,
causing unphysical numerical reflections at the boundary. In principle,
these reflections can be made arbitrarily small for $\triangle t\to 0$.
However, for practical time step sizes, the calculation of the current density
would be still distorted.
Our (new) idea is to apply a Crank-Nicolson discretization to a 
differential equation satisfied by $\exp(i\int_0^t V_L(s)ds/\hbar)$. 
Indeed, this expression solves
$$
  \frac{d\eps}{dt}(t) = \frac{i}{\hbar}V_L(t)\eps(t), \quad \eps(0)=1.
$$
The Crank-Nicolson discretization of this ordinary differential equation reads as
$$
  \eps^{(n+1)} = \eps^{(n)} + \triangle t\frac{i}{2\hbar} V_L^{(n+1/2)}(\eps^{(n+1)}+\eps^{(n)}),
  \quad \eps^{(0)} = 1.
$$
This recursion relation can be solved explicitly yielding
\begin{equation*}
  \eps^{(n)} = \exp\left(2i\sum_{\ell=0}^{n-1}\arctan\left(\frac{\triangle t}{2\hbar}
  V_L^{(\ell+1/2)}\right)\right), \quad n\in\N_0.
\end{equation*}
A Taylor series expansion
\[
2 i \arctan \left( \frac{\triangle t}{2 \hbar} V_L^{(\ell+ 1/2)} \right)
= 
\frac{i}{\hbar} V_L^{(\ell+1/2)} \triangle t + O((\triangle t)^3)
\]
reveals that in the limit $\triangle t \rightarrow 0$, the ad-hoc discretization in 
\eqref{eq.discrete_gauge_change_non_consistent} coincides with the discrete gauge change which is derived from the Crank-Nicolson time-integration method.

Analogously, $\exp(-i(E-V_L(0))t/\hbar)$ needs to be replaced by
\begin{align*}
  \gamma_J^{(n)} &:=
  \exp \left(2i \sum_{\ell=0}^{n-1}
  \arctan\left(-\frac{\triangle t}{2\hbar}E\right)\right)
  \exp \left(2i\sum_{\ell=0}^{n-1} 
  \arctan\left( \frac{\triangle t}{2 \hbar} V_L^{(0)} \right)\right) \\
  &= \exp\left[2in \left( \arctan\left(\frac{\triangle t}{2 \hbar} V_L^{(0)}\right) 
  - \arctan\left(\frac{\triangle t}{2 \hbar } E \right) \right)\right], \quad n\in\N_0.
\end{align*}
Thus, the discrete analogon of $\varphi$ in definition \eqref{3.phi} is given by
$$
  \varphi_j^{(n)} = \psi_j^{(n)} \eps^{(n)} - \phi_j\gamma_J^{(n)}, \quad
  j\in\{0,\ldots,J\}, \ n\in\N_0.
$$
Replacing $\psi_j^{(n)}$ by $\varphi_j^{(n)}$ in 
\eqref{eq.discrete_homogeneous_transparent_boundary_conditions_2},
we obtain the desired nonhomogeneous discrete transparent boundary condition
at $x_J=L$:
\begin{equation}\label{eq.nonhomogeneous_discrete_transparent_boundary_condition_right}
  \begin{aligned}
  \psi_{J-1}^{(n+1)} \epsilon^{(n+1)} - s^{(0)} \psi_J^{(n+1)} \epsilon^{(n+1)}
  &= -\epsilon^{(n)} \psi_{J-1}^{(n)} + \sum_{\ell=1}^{n} s^{(n+1-\ell)} 
  \left( \psi_J^{(\ell)} \epsilon^{(\ell)} - \phi_J \gamma_J^{(\ell)} \right) \\
  &\phantom{xx}{}- s^{(0)} \phi_J \gamma_J^{(n+1)}
  + \phi_{J-1} \left( \gamma_J^{(n+1)} + \gamma_J^{(n)} \right).
\end{aligned}
\end{equation}

At the left contact $x_0=0$, a nonhomogeneous boundary condition can be derived
in a similar way. Since the potential energy in the left lead is assumed to vanish,
the term $\eps^{(n)}$ is not needed, and the boundary condition is given by
\begin{equation}\label{eq.nonhomogeneous_discrete_transparent_boundary_condition_left}
  \begin{aligned}
  \psi_1^{(n+1)} - s^{(0)} \psi_0^{(n+1)} 
  &= - \psi_1^{(n)} + \sum_{\ell=1}^{n} s^{(n+1-\ell)} \left(\psi_0^{(\ell)} 
  - \phi_0\gamma_0^{(\ell)} \right) \\
  &\phantom{xx}{}- s^{(0)} \phi_0 \gamma_0^{(n+1)}
  + \phi_1 \left( \gamma_0^{(n+1)} + \gamma_0^{(n)} \right),
\end{aligned}
\end{equation} 
where
$$
  \gamma_0^{(n)} := \exp \left(-2in \arctan\left(\frac{\triangle t}{2 \hbar} E \right)
  \right), \quad  n \in \N_0.
$$

We summarize: The Crank-Nicolson scheme \eqref{3.cn} with the nonhomogeneous
discrete transparent boundary conditions 
\eqref{eq.nonhomogeneous_discrete_transparent_boundary_condition_right} 
and \eqref{eq.nonhomogeneous_discrete_transparent_boundary_condition_left}
reads as
\begin{equation}\label{eq.propagate_discrete_wave_function}
  B\psi^{(n+1)} = C\psi^{(n)} + d^{(n)},
\end{equation}
where $\psi^{(n)}=(\psi_0^{(n)},\ldots,\psi_J^{(n)})^\top$, $d=(d_0^{(n)},0,\ldots,0,d_J^{(n)})^\top$.
Furthermore, $B$ is a tridiagonal matrix with main diagonal
$(-s^{(0)},iR-2+wV_1^{(n+1/2)},\ldots,iR-2+wV_{J-1}^{(n+1/2)},-s^{(0)}\eps^{(n+1)})$,
upper diagonal $(1,\ldots,1)$, and lower diagonal $(1,\ldots,1,\eps^{(n+1)})$;
$C$ is a tridiagonal matrix with main diagonal
$(0,iR+2-wV_1^{(n+1/2)},\ldots,iR+2-wV_{J-1}^{(n+1/2)},0)$,
upper diagonal $(-1,\ldots,-1)$ and lower diagonal $(-1,\ldots,-1,-\eps^{(n)})$;
furthermore,
\begin{align}
  d_0^{(n)} &= 
  \sum_{\ell=1}^{n} s^{(n+1-\ell)} \left( \psi_0^{(\ell)} - \phi_0 \gamma_0^{(\ell)} \right)
  - s^{(0)} \phi_0 \gamma_0^{(n+1)} + \phi_1 \left(\gamma_0^{(n+1)} + \gamma_0^{(n)}\right), 
  \label{eq.propagate_discrete_wave_function_inhomogeneity_lhs} \\
  d_J^{(n)} &= 
  \sum_{\ell=1}^n s^{(n+1-\ell)} \left( \psi_J^{(\ell)} \epsilon^{(\ell)} 
  - \phi_J \gamma_J^{(\ell)} \right) - s^{(0)} \phi_J \gamma_J^{(n+1)} 
  + \phi_{J-1} \left( \gamma_J^{(n+1)} + \gamma_J^{(n)} \right).
\label{eq.propagate_discrete_wave_function_inhomogeneity_rhs}
\end{align}


\subsection{Fast evaluation of the discrete convolution terms}\label{sec.trans.conv}

In the subsequent simulations, scheme \eqref{eq.propagate_discrete_wave_function}
has to be solved in each time step and for every wave function
$\psi=\psi_k$, $k\in{\mathcal K}$.
We recall that the kernel coefficients $s^{(n)}$ need to be calculated 
only once as they do not depend on the wave number $k$.
Let $N$ denote the number of time steps.
For each $k\in{\mathcal K}$,
we require order $O(N)$ storage units and $O(N^2)$ work units to compute the
discrete convolutions in 
\eqref{eq.propagate_discrete_wave_function_inhomogeneity_lhs} and
\eqref{eq.propagate_discrete_wave_function_inhomogeneity_rhs}.
For this reason, simulations with several ten thousands of time steps are not
feasible. To overcome this problem, one may truncate the convolutions at some
index, since the decay rate of the convolution coefficients is of order
$O(n^{-3/2})$ \cite[Section 3.3]{EhAr01}. The drawback of this approach is that still
more than thousand convolution terms are necessary to avoid unphysical reflections at the boundaries. 

The problem has been overcome in \cite{AES03} by approximating
the original convolution coefficients $s^{(n)}$ and calculating 
the approximated convolutions
by recursion. More precisely, approximate $s^{(n)}$ by
$$
  \widetilde s^{(n)} := 
  \begin{cases}
  s^{(n)},                        \quad &n < \nu, \\ 
  \sum_{\ell = 1}^\Lambda b_l q_l^{-n}, \quad &n \geq \nu,
  \end{cases}
$$
such that
\begin{equation}\label{eq.approximate_convolution}
  \mathcal{C}^{(n)}(u) := \sum_{\ell = 1}^{n-\nu} \widetilde{s}^{(n-\ell)} u^{(\ell)} 
  \approx \sum_{\ell = 1}^{n-\nu} s^{(n-\ell)} u^{(\ell)}
\end{equation}
can be evaluated by a recurrence formula which reduces the numerical effort
drastically. As in \cite{AES03}, we set $\nu=2$ to exclude $s^{(0)}$ and
$s^{(1)}$ from the approximation. In fact, $s^{(0)}$ does not appear in the
original convolutions, whereas $s^{(1)}$ is excluded to increase the accuracy.

Let $\Lambda\in\N$. The set $\{b_0,q_0\ldots,b_\Lambda,q_\Lambda\}$ 
is computed as follows. First, define the formal power series
$$
  h(x) := s^{(\nu)} + s^{(\nu+1)} x + s^{(\nu+2)} x^2 + \cdots 
  + s^{(\nu+2\Lambda-1)} x^{\nu+2\Lambda-1} + \cdots, \quad |x| \leq 1.
$$
The first (at least $2\Lambda$) coefficients are required to calculate the
$[\Lambda-1|\Lambda]$-Pad\'e approximation of $h$, $\widetilde 
h(x):=P_{\Lambda-1}(x)/Q_\Lambda(x)$,
where $P_{\Lambda-1}$ and $Q_\Lambda$ 
are polynomials of degree $\Lambda-1$ and $\Lambda$, respectively.
If this approximation exists, we can compute its Taylor series
$\widetilde h(x)=\widetilde s^{(\nu)}+\widetilde s^{(\nu+1)}x+\cdots$,
and by definition of the Pad\'e approximation, it holds that
$$
  \tilde{s}^{(n)} = s^{(n)} \quad\textrm{for all }n\in
  \{\nu,\nu+1,\ldots,\nu+2\Lambda-1\}.
$$
It can be shown that, if $Q_\Lambda$ has $\Lambda$ simple roots $q_\ell$ with
$|q_\ell|>1$ for all $\ell\in\{1,\ldots,\Lambda\}$, the approximated coefficients
are given by
\begin{equation}\label{eq.s_tilde}
  \tilde{s}^{(n)} = \sum_{\ell = 1}^\Lambda b_\ell q_l^{-n}, \quad 
  b_\ell := - \frac{P_{\Lambda-1}(q_\ell)}{Q_\Lambda(q_\ell)} 
  q_\ell^{\nu-1} \neq 0, \quad
  n\ge \nu,\ \ell\in\{1,\ldots,\Lambda\}.
\end{equation}

Summarizing, one first computes the exact coefficients 
$s^{(0)},\ldots,s^{(\nu+2\Lambda-1)}$
followed by the $[\Lambda-1|\Lambda]$-Pad\'e approximation. 
Then one determines the roots
of $Q_\Lambda$, yielding the numbers $q_1,\ldots,q_\Lambda$. 
Finally, one evaluates \eqref{eq.s_tilde}
to find the coefficients $b_0,\ldots,b_\Lambda$. We stress the fact that these
calculations have to be performed with high precision ($2\Lambda-1$ mantissa length)
since otherwise the Pad\'e approximation may fail (see \cite{AES03}). 
We employ the Python library {\tt mpmath} for arbitrary-precision
floating-point arithmetics \cite{Joh10}. As an alternative, one may use the 
{\tt Maple} script from \cite[Appendix]{AES03}. 

A particular feature of this approximation is that it can be calculated by
recursion. More precisely, for $n\ge\nu+1$, the function ${\mathcal C}^{(n)}(u)$
in \eqref{eq.approximate_convolution} can be written as
$$
  \mathcal{C}^{(n)}(u) = \sum_{\ell=1}^{\Lambda} C_\ell^{(n)}(u),
$$
with
$$
  C_\ell^{(n)}(u) = q_\ell^{-1} C_\ell^{(n-1)}(u) + b_\ell q_\ell^{-2} u^{(n-2)},\quad 
  n \geq \nu + 1, \quad C_\ell^{(\nu)}(u) = 0.
$$
Hence, the discrete convolutions in
\eqref{eq.propagate_discrete_wave_function_inhomogeneity_lhs} and
\eqref{eq.propagate_discrete_wave_function_inhomogeneity_rhs} are approximated
for $n\ge\nu=2$ by
\begin{equation}
\label{eq.approximated_convolution}
  \sum_{\ell=1}^n s^{(n+1-\ell)} u^{(\ell)}
  \approx \mathcal{C}^{(n+1)} \left( u \right) + s^{(1)} u^{(n)},
\end{equation}
whereas the exact expressions are used for $n=0$ and $n=1$. 
As a result, the storage for the implementation of the discrete transparent boundary
conditions reduces from $O(N)$ to $O(\Lambda)$. Even more importantly, the work
is of order $O(\Lambda N)$ instead of $O(N^2)$. 

Obviously, the quality of the approximation depends on $\Lambda$. 
By construction, we have $s^{(n)}=\widetilde s^{(n)}$ for all $n\in\{0,\ldots,2\Lambda+\nu-1\}$ 
but $\widetilde s^{(n)}$ approximates $s^{(n)}$ very well even if $n$ is much larger \cite{AES03}.
We illustrate in Section \ref{section_convergence_in_lambda} that
the convergence of the complete transient algorithm with respect to $\Lambda$
is exponential.


\subsection{The complete transient algorithm}\label{sec.trans.alg}

In the previous sections, we have explained the approximation of the
transient Schr\"o\-din\-ger equation with discrete transparent 
boundary conditions for
given potential energy $V=V_{\rm barr}+V_{\rm self}$. Here, we make explicit the
coupling procedure with the Poisson equation for the selfconsistent
potential
$$
  -\frac{\pa^2 V_{\rm self}}{\pa x^2} = \frac{e^2}{\eps}(n[V_{\rm self}]-n_D),
  \quad x\in(0,L), \quad V_{\rm self}(0,t)=0, \ V_{\rm self}(L,t)=-eU(t),
$$
with the electron density
$$
  n[V_{\mathrm{self}}](x,t) = \int_{\R} g(k) |\psi_k(x,t)|^2 \,dk.
$$
According to the Crank-Nicolson scheme, a natural approach would be to
employ a two-step predictor-corrector scheme. More precisely, let
$\{\psi_k^{(n)}\}_{k\in{\mathcal K}}\to\{\psi_k^{(*)}\}_{k\in{\mathcal K}}$ be
propagated for one time step using $V_{\rm self}^{(n)}$ to obtain
$V_{\rm self}^{(*)}$. Then one uses
$V_{\rm self}^{(n+1/2)}:=\frac12(V_{\rm self}^{(n)}+V_{\rm self}^{(*)})$ to
propagate $\{\psi_k^{(n)}\}_{k\in{\mathcal K}}\to\{\psi_k^{(n+1)}\}_{k\in{\mathcal K}}$
again. This procedure doubles the numerical effort and is computationally too
costly. As an alternative, the scheme
$V_{\rm self}^{(n+1/2)} := 2V_{\rm self}^{(n)} - V_{\rm self}^{(n-1/2)}$
can be employed (as in \cite{Pin02}).
We found in our simulations that the most simple approach, 
$V_{\rm self}^{(n+1/2)}:=V_{\rm self}^{(n)}$, gives essentially the same results
as the above schemes. 
The reason is that the electron density evolves
very slowly compared to the small time step size which is needed to resolve the fast oscillations of the wave functions.
Hence, the variations of $V_{\rm self}$ are small.
Similarly, the right boundary condition of the Poisson equation can be replaced 
by $V_{\mathrm{self}}(L) = -e U(n \triangle t)$ if the applied voltage varies slowly.
This is used in the circuit simulations of Section \ref{sec.circuit}.

\begin{figure}[htb]
\begin{centering}
\includegraphics[width=150mm]{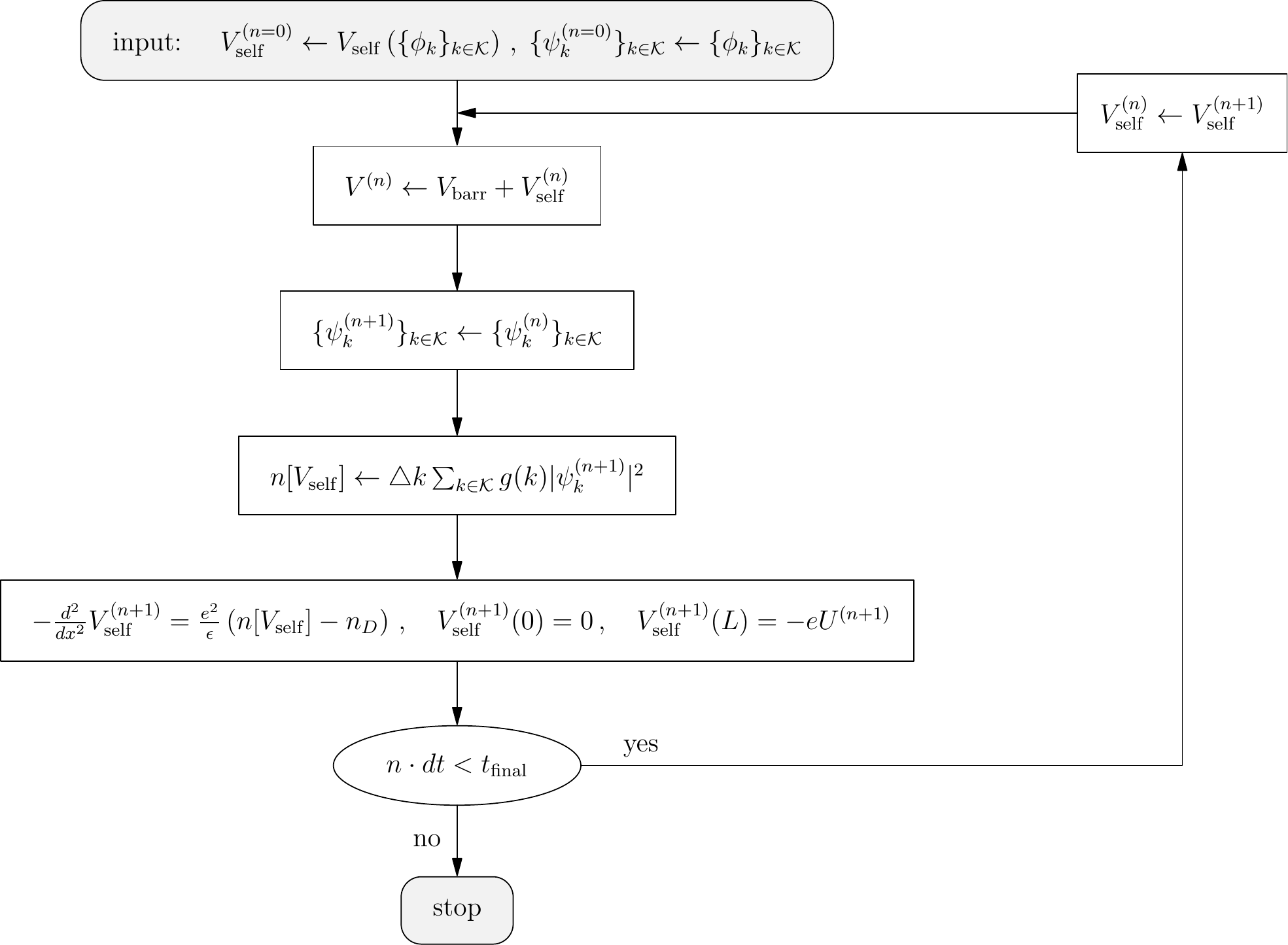}
\caption{Flow chart of the transient scheme.}
\label{fig:transient_algorithm}
\end{centering}
\end{figure}

The complete transient algorithm is presented in Figure \ref{fig:transient_algorithm}.

\subsection{Discretization parameters}

We choose $K=3000$ for the number of wave functions as in the
stationary simulations and $\triangle t=1$\,fs (fs = femtosecond) 
for the time step size.
With the maximal kinetic energy of injected electrons $\hbar\omega_M = \hbar^2 k_M^2/(2m)$, where $k_M$ is the maximal wave number, 
the period is computed according to $\tau_M=2\pi/\omega_M$. 
Thus, the fastest wave oscillation
is resolved by $\tau_M/\triangle t\approx 18.5$ time steps.
The space grid size is chosen to be $\triangle x=0.1$\,nm.
Consequently, the smallest wave length $\lambda_M = 2 \pi / k_M \approx 10$\,nm 
is resolved by approximately 100 spatial grid points. 
Furthermore, we take $\Lambda=70$ for the approximation parameter
of the discrete convolution terms.
This choice results from a numerical convergence study presented in Section
\ref{section_convergence_in_lambda}.

It is important to note that the wave functions which are propagated using 
the fast evaluation of the approximated discrete convolution terms \eqref{eq.approximated_convolution}
practically coincide with the wave functions which are propagated using the exact 
convolutions \eqref{eq.propagate_discrete_wave_function_inhomogeneity_lhs}--\eqref{eq.propagate_discrete_wave_function_inhomogeneity_rhs}
(see Section \ref{section_convergence_in_lambda}).
Employing the exact convolutions, however, is equivalent to solving the 
Crank-Nicolson finite difference equations of the whole space problem.
Considering that the electron density evolves smoothly in space and time, it is clear that 
the error of the complete transient algorithm (see Section \ref{sec.trans.alg}) is determined by
the Crank-Nicolson finite difference scheme.
A global error estimate, together with a meshing strategy depending on a 
possibly scaled Planck constant $\hbar$ is given in \cite{BaJiMa02}.
The calculations in this article are performed in SI units without any scaling.


\subsection{Details of the implementation}

The final solver is implemented in the {\tt C++} programming language
using the matrix library {\tt Eigen} \cite{Gue10} for concise and efficient
computations. 
As we are interested in simulations
with a very large number of time steps $N$ (e.g., $N=100\,000$),
some sort of parallelization is indispensable. We employ the library
{\tt pthreads} to realize multiple threads on multi-core processors with
shared memory. The most time consuming part in the transient algorithm
(see Section \ref{sec.trans.alg}) is the propagation of the wave functions
and the calculation of the electron density. Since the wave functions
evolve independently of each other, this task can be easily parallelized.
At every time step, we create a certain number of threads (usually, this number
equals the number of cores available). To each thread, we assign a subset of
wave functions which are propagated as described above. Before the threads are
joined again, each thread computes its part of the electron density.
All these parts provide the total eletron density which is used to solve
the Poisson equation in serial mode. The simulations presented below have been
carried out on an Intel Core 2 Quad CPU Q9950 with
$4\times 2.8$\,GHz.


\section{Numerical experiments}\label{sec.numer}

We present three numerical examples. The first example shows the importance to
provide a complete compatible discretization of the open Schr\"odinger-Poisson system.
The second numerical test shows the time-dependent behavior of a resonant
tunneling diode, which allows us to identify three physical regions.
In the third experiment, we investigate the convergence of our solver with 
respect to the parameter $\Lambda$ which appears in the context of the fast evaluation
of the discrete convolution terms.


\subsection{First experiment: Constant applied voltage}\label{sec.numer.1}

We compute the
stationary solution to the Schr\"odinger-Poisson system with an applied voltage
of $U=250$\,mV. At this voltage, the current density
achieves its first local maximum. 
We apply the transient algorithm of Section \ref{sec.trans} until $t=25$\,fs, 
keeping the applied voltage constant.
Accordingly, the stationary solution should be preserved
and the current density $J_{\mathrm{cond}}$,
defined in \eqref{2.J}, is expected to be spatially constant.

\begin{figure}[ht]
\begin{centering}
\includegraphics[width=150mm]{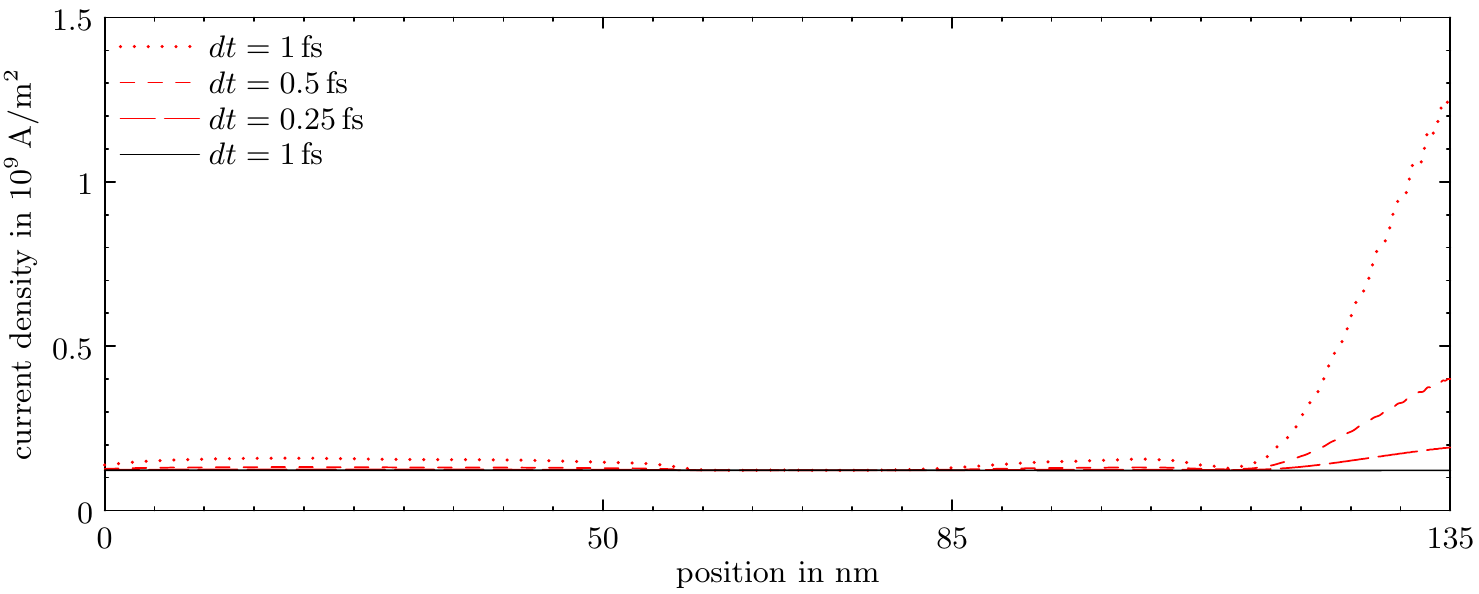}
\caption{
Conduction current density in a resonant tunneling diode at $t=25$\,fs for 
a constant applied voltage of $U=250$\,mV.
Discretizations using the ad-hoc discretization \eqref{eq.discrete_gauge_change_non_consistent}
of the analytical boundary conditions yield strongly distorted numerical 
solutions (broken lines).
In contrast, the conduction current density computed with our solver is perfectly constant (solid line).}
\label{fig.consistent_vs_inconsistent_discretization}
\end{centering}
\end{figure}

The ad-hoc discretization \eqref{eq.discrete_gauge_change_non_consistent}
is employed using the time step sizes $\triangle t=1$\,fs, 0.5\,fs, 0.25\,fs. 
We observe in
Figure \ref{fig.consistent_vs_inconsistent_discretization} that the current
density is not constant. The reason is that the discretization
\eqref{eq.discrete_gauge_change_non_consistent} is not compatible with the
underlying finite-difference scheme. The distortions are reduced for very small
time step sizes but this leads to computationally expensive algorithms. 
In contrast, with the discrete gauge change of Section \ref{sec.trans.dtbc},
the current density is perfectly constant even for the rather large time step
$\triangle t=1$\,fs; see Figure \ref{fig.consistent_vs_inconsistent_discretization}.

We mention that the transient solution is also distorted if the scattering states
as initial wave functions are computed from an ad-hoc discretization of the
continuous boundary conditions \eqref{eq.analytic_stationary_boundary_conditions_1}
and \eqref{eq.analytic_stationary_boundary_conditions_2}. For
stationary computations, spurious reflections due to an inconsistent
discretization play a minor role but they become a major issue for transient
simulations.


\subsection{Second experiment: Time-dependent applied voltage}\label{sec.numer.2}

For the second numerical experiment, we consider a time-dependent applied voltage.
The conduction current density is no longer constant but 
the total current density is expected to be conserved.
We recall that the total current density $J_{\mathrm{tot}} = J_{\mathrm{cond}} + \pa D/\pa t$
is the sum of the conduction current density $J_{\mathrm{cond}}$ and the displacement current density $\pa D/\pa t$.
Here $D$ denotes the electric displacement field which is related to the electric field $E$ by $D = \epsilon_0 \epsilon_r E$.
Indeed, replacing the electric field by the negative gradient of the potential we obtain
\[
  \frac{\partial D}{\partial t} 
  = -\frac{\epsilon_0 \epsilon_r}{e} \frac{\partial}{\partial t} \na V_{\mathrm{self}}.
\]
The temporal and spatial derivatives are approximated using centered finite differences.
Amp\`ere's circuital law
$\na\times H=J_{\rm tot}$ for the magnetic field strength $H$ yields
$$ 
  \diver\, J_{\rm tot} = \diver(\na\times H) = 0, 
$$
and hence, in one space dimension, $J_{\rm tot}$ is constant in space.

\begin{figure}
\begin{center}
\includegraphics[width=65mm]{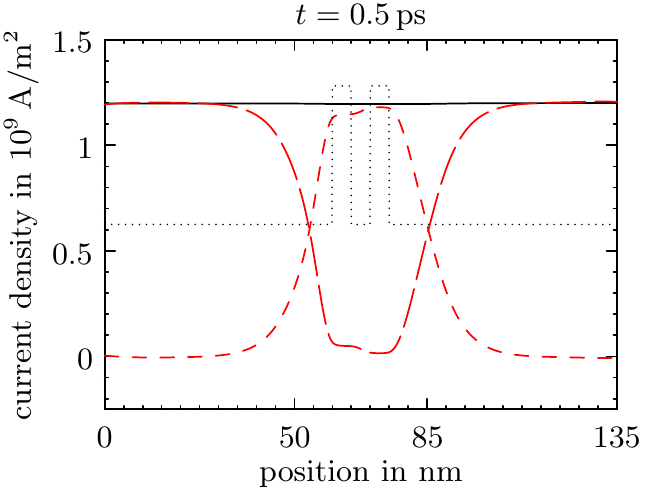}
\quad \includegraphics[width=65mm]{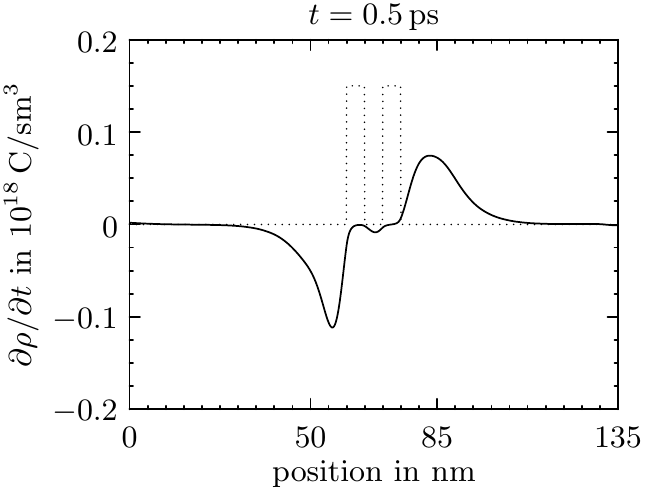}
\\
\vspace{2.5mm}
\includegraphics[width=65mm]{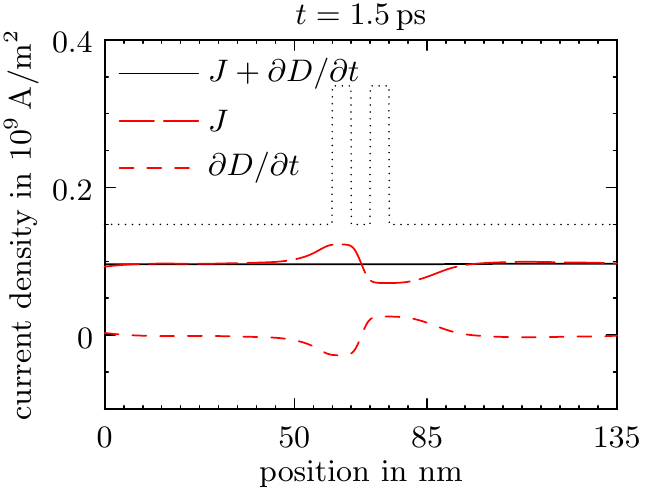}
\quad \includegraphics[width=65mm]{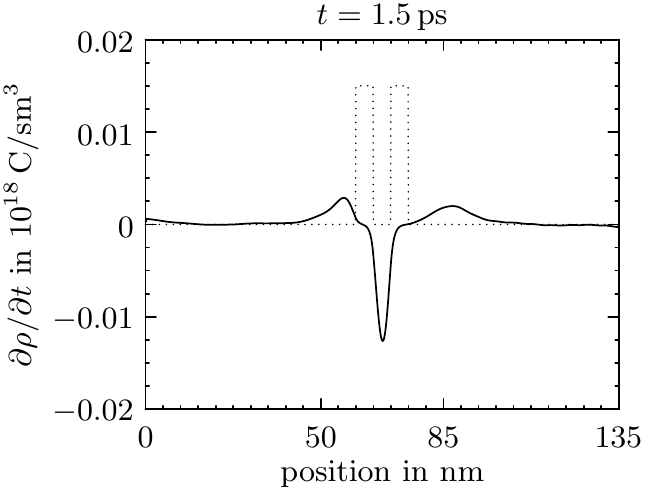}
\\
\vspace{2.5mm}
\includegraphics[width=65mm]{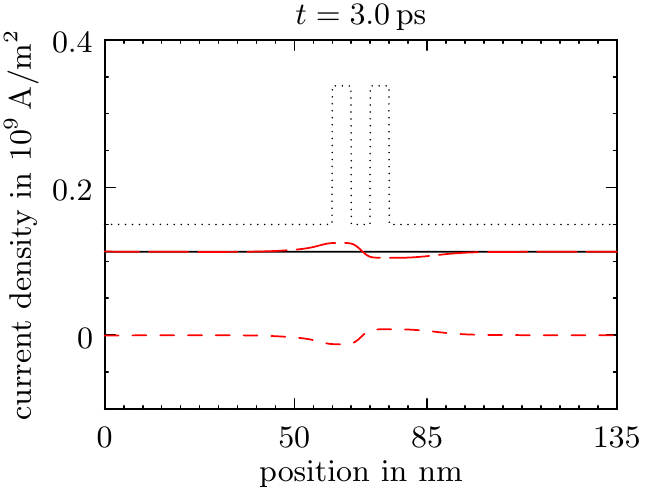}
\quad \includegraphics[width=65mm]{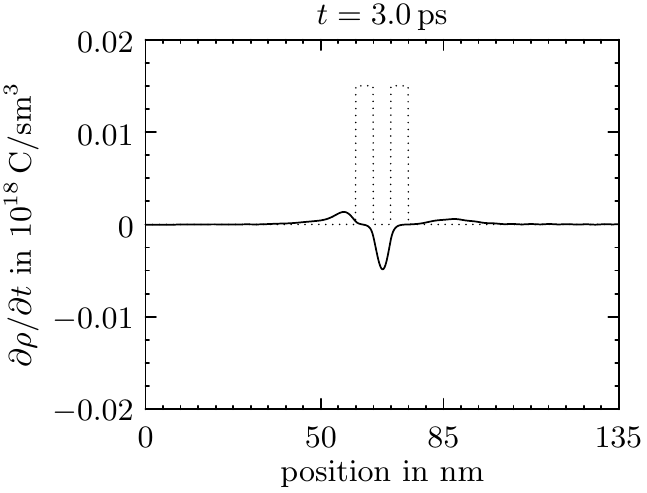}
\end{center}
\caption{Left column: Total current density $J_{\rm tot}=J_{\rm cond}+\pa D/\pa t$, 
conduction current density $J:=J_{\rm cond}$, 
and displacement current density $\pa D/\pa t$ versus position
at different times.
Right column: Temporal variation $\pa \rho/\pa t$ of the charge density 
versus position.}
\label{fig:rtd_current_density_at_different_times}
\end{figure}

The following simulation demonstrates that the total current density is a conserved quantity in the discrete system as well.
First, we compute the equilibrium state using an applied voltage of $U=0$\,V. 
This solution is then propagated using a raised cosine function for the applied voltage
$$
  U(t) = \frac{U_0}{2}\left(1-\cos\frac{2\pi t}{T}\right), 
  \quad 0\le t\le 1\,\mbox{ps},
$$
where $U_0=0.25$\,V and $T=2$\,ps. At later times, $t\ge 1$\,ps, $U(t)=U_0$ is kept
constant.
Conduction, displacement, and total current density at different times
are depicted in the left column of Figure \ref{fig:rtd_current_density_at_different_times}.
As can be seen, the total current density is perfectly conserved at all
considered times. 
The change of the charge density $\pa \rho/\pa t$ is illustrated in the right column of Figure \ref{fig:rtd_current_density_at_different_times}.
In our model, $\rho$ is given by $\rho = e \left( n_D - n\right)$.

\begin{figure}[t]
\centering
\includegraphics[width=150mm]{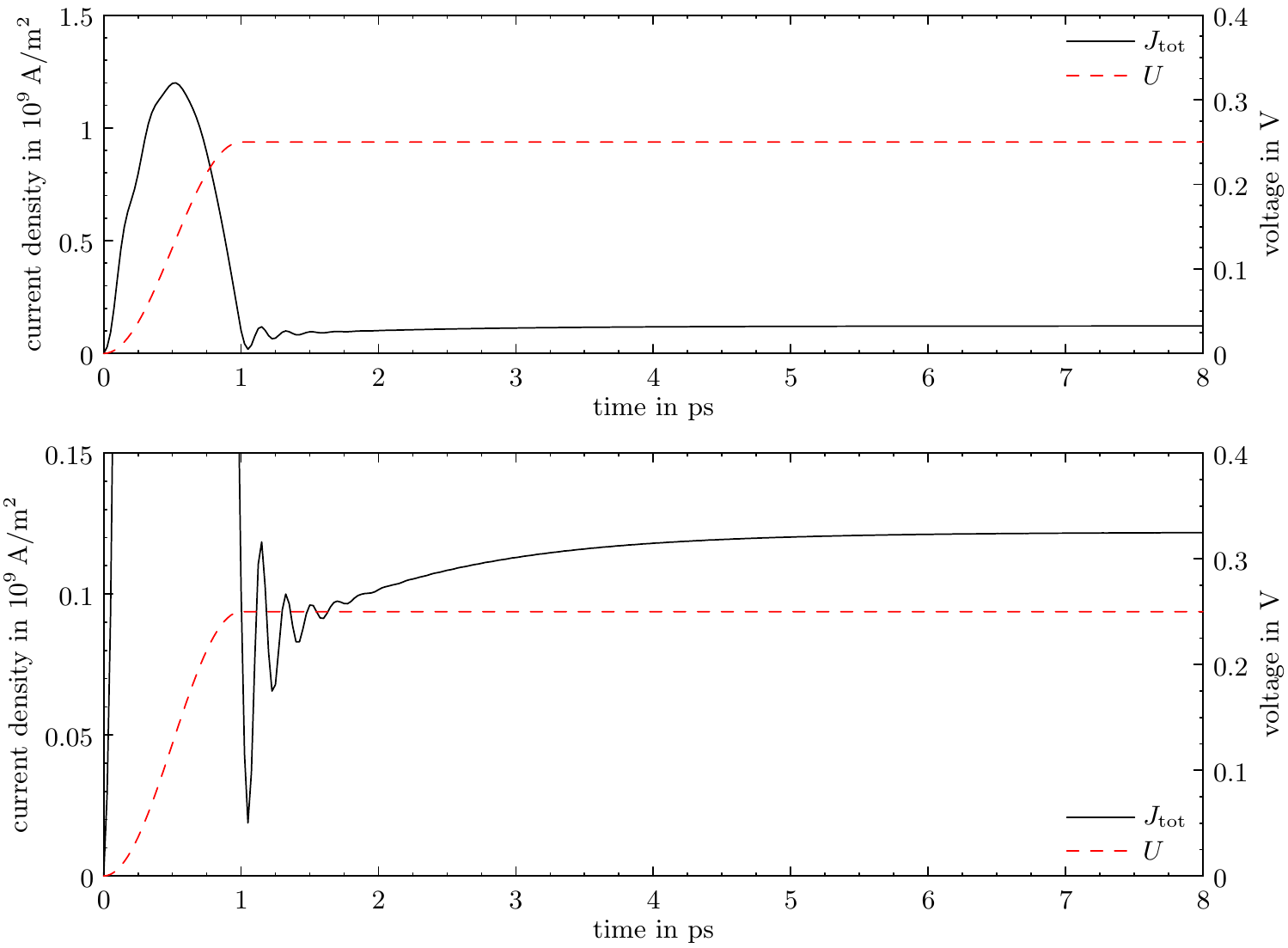}
\caption{
Applied voltage and total current density versus time in different scalings.}
\label{fig:rtd_applied_voltage_and_current_density}
\end{figure}

The time-dependence of the total current density in response to the applied voltage 
is shown in Fig. \ref{fig:rtd_applied_voltage_and_current_density}. 
We can identify three different regions in the temporal behaviour, each of which is governed by a different physical mechanism.

\medskip\noindent
{\em Region I: Capacitive behavior.} When the applied voltage increases during
the first picosecond, the resonant tunneling diode behaves mainly like a 
parallel plate capacitor. This can be clearly seen in the top left panel of
Figure \ref{fig:rtd_current_density_at_different_times}. In the region of
the double barrier, the displacement current gives the dominant contribution
to the total current, whereas the conduction current is small. The top right panel
of Figure \ref{fig:rtd_current_density_at_different_times} shows a
build-up of negative charge before the left barrier and of positive charge
after the right barrier. The formation of opposite charges on the two sides
of the double barrier results in the formation of an electric field between the
two regions of opposite charge density. This field is necessary to accomodate
the externally applied voltage. Figure \ref{fig:rtd_applied_voltage_and_current_density}
shows that the current closely follows the time derivative of the applied voltage:
$$
  J_{\rm cond} \approx 
  C \frac{dU}{dt} = \frac{\pi C U_0 }{T}\sin\left(\frac{2\pi t}{T}\right).
$$
This expression allows us to estimate the apparent capacitance $C$. The maximum
current density occurring at $t = T/4 = 0.5$\,ps takes approximately the value
$1.2\cdot 10^9$\,Am$^{-2}$. We compute
$C = T J / \pi U_0 =3.06\cdot 10^{-3}$\,Fm$^{-2}$. Equating this value to the parallel plate capacitance, $C = \eps_0 \eps_r/ d$, we
find the average separation of the opposite charge densities to be $d = 33.1\,$nm. 

\medskip\noindent
{\em Region II: Plasma oscillations.}
During the second picosecond, a strongly damped oscillation occurs in the current density. From Figure \ref{fig:rtd_applied_voltage_and_current_density}, we estimate
five oscillations to occur during one picosecond, which relates to a period of 
about 200\,fs. 
It is believed that these are plasma oscillations which were excited by the rapidly changing applied voltage $U$. As soon as the transient phase of $U(t)$ is over and
$U(t)$ is kept constant at $U_0$ for $t \ge 1$\,ps, the excitation vanishes and the oscillations  fade out quickly. As a rough estimate we calculate the plasma frequency
$\omega_p$ for a classical electron system of uniform density:
$$
  \omega_p^2 = \frac{n e^2}{m \eps_0}.
$$
Note that in the resonant tunneling diode the density is neither uniform nor is it governed by the classical equations of motion. Nevertheless, we may use
this expression to estimate the order of magnitude of the time constant associated with this effect. Since plasma oscillations
usually occur in the high-density regions of a  device, we set $n=n_D^1 = 10^{24}$\,m$^{-3}$ 
and obtain $\tau_p = 2\pi/\omega_p = 111.4$\,fs. 
This value is of the same order 
as the $200$\,ps estimated above, which is a strong indication that the physical effect observed here is a plasma oscillation.

\begin{figure}
\centering
\includegraphics[width=150mm]{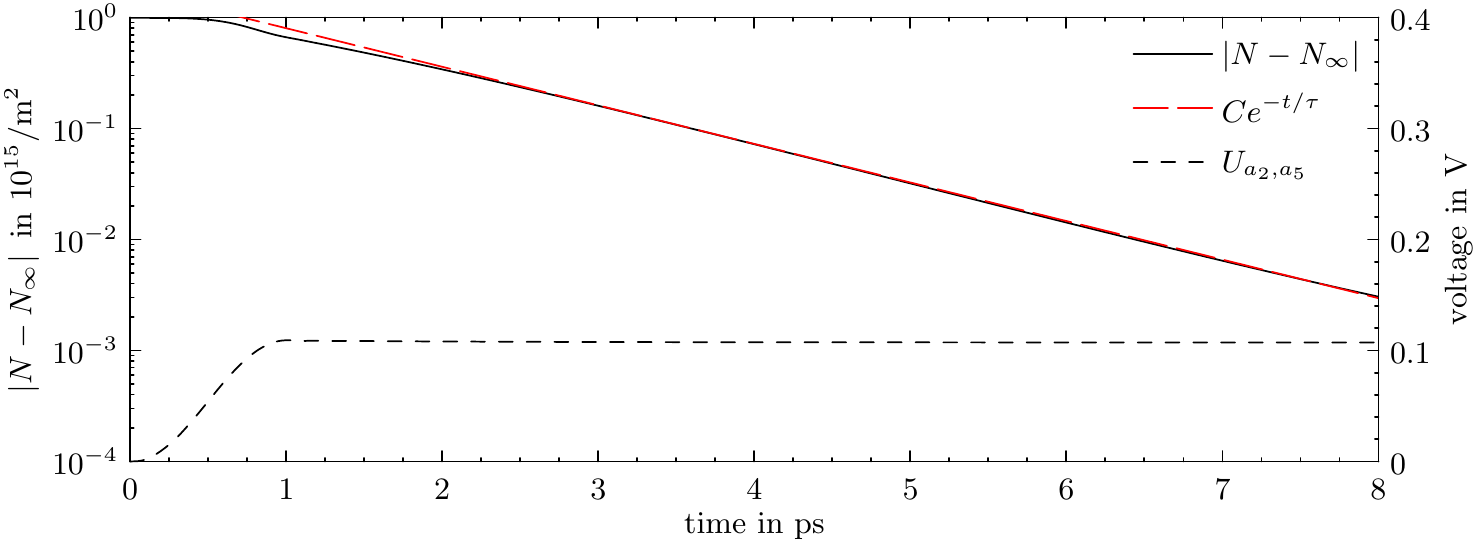}
\caption{Number of electrons in the quantum well versus time. 
In Region III ($t \ge 2$\,ps) this number clearly follows an exponential law.
$U_{a_2,a_5}$ denotes the temporal variation of the voltage between $x=a_2$ and $x=a_5$.}
\label{fig:tau}
\end{figure}

\medskip\noindent
{\em Region III: Charging of the quantum well.} 
For $t > 2$\,ps, an exponential increase in the current can be clearly observed in 
Figure \ref{fig:rtd_applied_voltage_and_current_density}. 
Below 2\,ps we see a superposition of both the exponential current increase and the plasma oscillations. 
The origin of this effect can be understood from the right panels of
Figure \ref{fig:rtd_current_density_at_different_times}.
Negative charge builds up in the quantum well. 
This charge  results from electrons tunneling through the left barrier into the quantum well. In this context, we note that the temporal variation of the voltage between 
the left and right end points $a_2$ and $a_5$ of the double-barrier structure,
respectively,
follows closely the variation of the applied voltage $U$ and hence, it is
practically constant for $t>1$\,ps (see Figure \ref{fig:tau}).
The rate $|\partial\rho/\partial t|$ decreases with time as can be seen by the snapshots at $t=1.5$\,ps and $t=3$\,ps.
We calculate the number of electrons residing in the quantum well:
$$
  N(t) := \int_{a_2}^{a_5} n(x,t) \, dx.
$$
Since the charging process is expected to show an exponential time dependence, we assume
the following exponential law for $N(t)$ and extract the free parameters $\tau$ and $N_\infty$:
$$
  N(t) = N_{\infty} + \left(N(t_1) - N_{\infty}\right)e^{-(t-t_1)/\tau}. 
$$
In Figure \ref{fig:tau}, the difference $|N(t) - N_\infty|$ is plotted, 
which decays to zero with an extracted time constant of $\tau = 1.25$\,ps. 

\begin{figure}
\includegraphics[width=75mm]{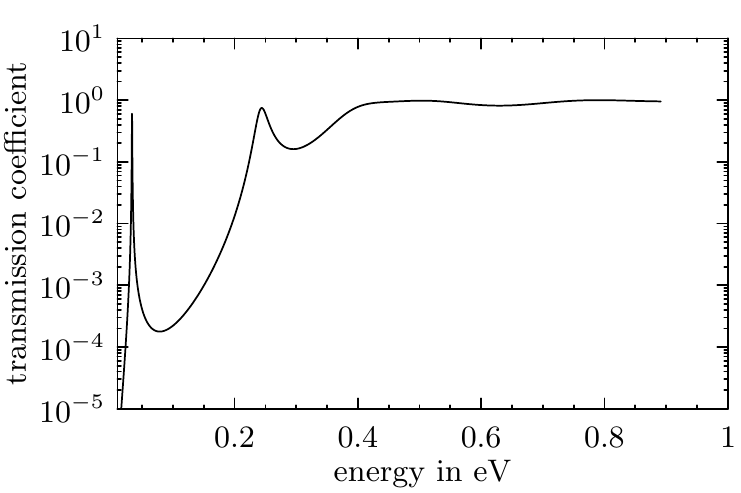}
\quad
\includegraphics[width=75mm]{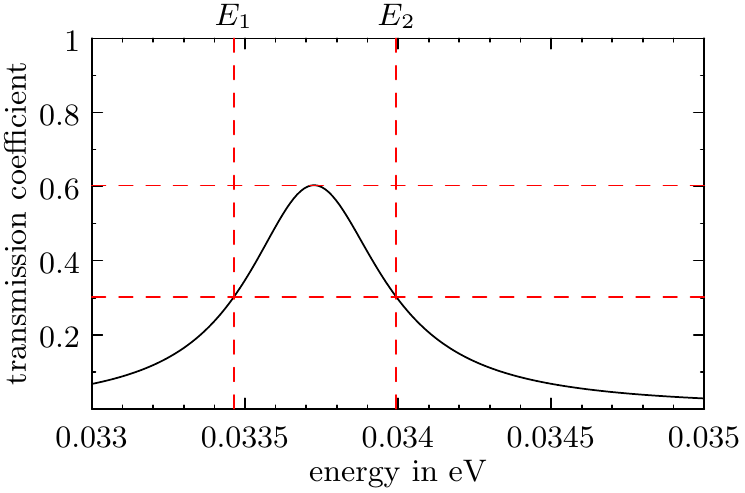}
\caption{
Transmission coefficient of the double-barrier structure at $U=0.25$\,V in different scalings.}
\label{fig:transmission}
\end{figure}

This time scale is related to the life time of a quasi-bound state. 
At $U=0.25$\,V, the current-voltage characteristic has its first maximum, which means that the first 
resonant state in the quantum well is carrying the current. 
The life time of this resonant state can be extracted from the width of the resonance peak in the transmission coefficient. 
Figure \ref{fig:transmission} depicts the transmission coefficient of the double-barrier structure 
at $U=0.25$\,V.
The transmission coefficient is defined as the ratio between the transmitted and the incident probability current density $j_{\mathrm{trans}}$
and $j_{\mathrm{inc}}$. In terms of the amplitude and the wavenumber of the transmitted and the incident wave, it reads: 
$$
  \frac{|j_{\mathrm{trans}}|}{|j_{\mathrm{inc}}|} 
  = \frac{|A_{\mathrm{trans}}|^2 k_{\mathrm{trans}}}{|A_{\mathrm{inc}}|^2
  k_{\mathrm{inc}}}.
$$
Extracting $\triangle E$, the half width at half maximum of the first transmission peak,
the life time of the resonant state can be estimated as follows \cite{KMR98}:
\begin{equation*}
  \tau = \frac{\hbar}{2 \triangle E}.
\end{equation*}
At $U=0.25$\,V we find $2\triangle E = 5.31\cdot 10^{-4}$\,eV and thus 
$\tau = 1.24$\,ps. This value is very close to the time constant of 
$\tau = 1.25$\,ps extracted from the exponential
charge increase in the quantum well, which is the cause for the observed
exponential current increase.


\subsection{Third experiment: Convergence in $\Lambda$}
\label{section_convergence_in_lambda}

\begin{figure}
\includegraphics[width=75mm]{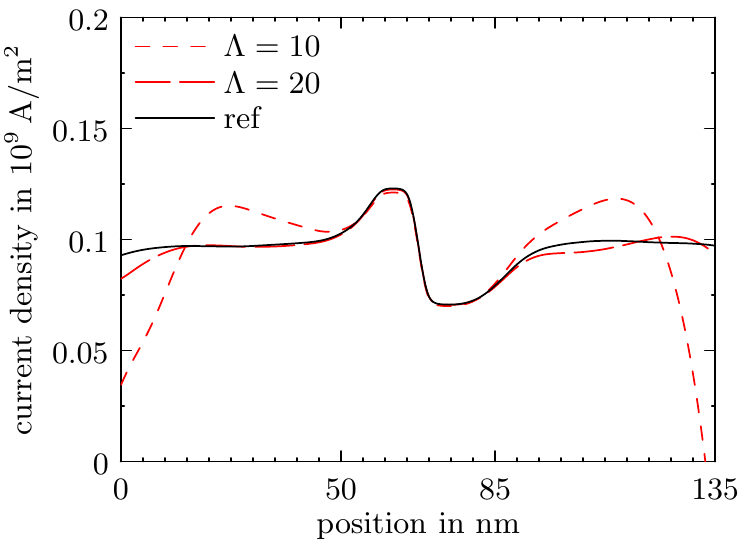}
\quad
\includegraphics[width=75mm]{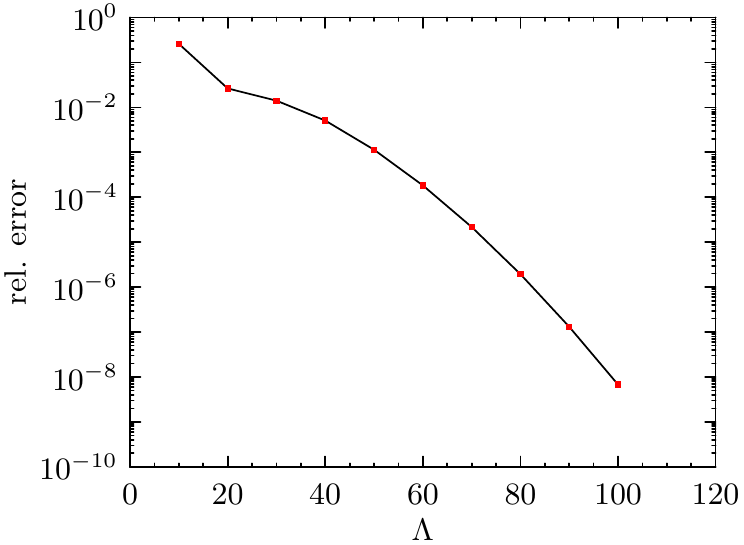}
\caption{
Conduction current density at $t=1.5$\,ps (left) and relative $\ell^2$-error for increasing $\Lambda$.}
\label{fig:convergence_in_Lambda}
\end{figure}

Finally, we study the convergence of the complete transient
algorithm detailed in Section \ref{sec.trans.alg} 
with respect to the parameter $\Lambda$ which appears in the context of the fast 
evaluation of the discrete convolution terms.
For this purpose, we repeat the last experiment with different values of $\Lambda$.
We compare the results with those obtained from the algorithm which uses 
the discrete transparent boundary conditions with the exact
convolutions \eqref{eq.propagate_discrete_wave_function_inhomogeneity_lhs}--\eqref{eq.propagate_discrete_wave_function_inhomogeneity_rhs}.
Since the computation of the reference solution is extremely expensive,
we restrict the experiment to the final time $t=1.5$\,ps.
The conduction current densities at $t=1.5$\,ps for two different values of $\Lambda$ 
and for the reference solution are depicted in 
Figure~\ref{fig:convergence_in_Lambda} (left). 
The relative error in the $\ell^2$-norm for increasing values of $\Lambda$ is 
shown in Figure~\ref{fig:convergence_in_Lambda} (right). 
We observe that the relative error decreases exponentially fast.
Thus, using a relatively small value of $\Lambda$ practically yields the same results
(at dramatically reduced numerical costs)
as if the discrete transparent boundary conditions with the exact convolutions
were used.


\section{Circuit simulations}\label{sec.circuit}

\begin{figure}[htb]
\begin{centering}
\includegraphics[width=75mm]{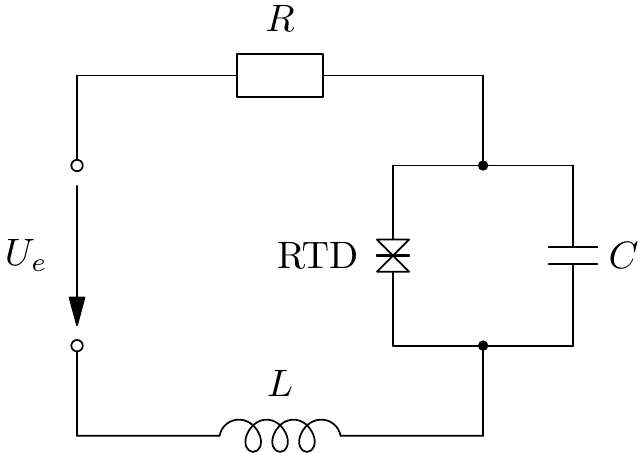}
\caption{High-frequency oscillator containing the resonant tunneling diode RTD.}
\label{fig.oscillator_circuit_layout}
\end{centering}
\end{figure}

In this section, we simulate a high-frequency oscillator consisting of a voltage
source $U_e$, a resistor with resistance $R$, an inductor with inductance $L$,
a capacitor with capacity $C$, and a resonant tunneling diode RTD; 
see Figure \ref{fig.oscillator_circuit_layout}. Each element of the circuit yields
one current-voltage relationship,
\begin{equation}\label{eq:current_voltage_relationships}
  U_R = RI_R, \quad U_L = L\dot I_L, \quad I_C = C\dot U_C, \quad
  I_{\rm RTD} = f(U_{\rm RTD}).
\end{equation}
The last expression is to be understood as follows. Given the applied voltage
$U_{\rm RTD}$ at the tunneling diode, the current $I_{\rm RTD}(t)=AJ_{\rm tot}(t)$
is computed from the solution of the time-dependent Schr\"odinger-Poisson system.
Here, $A=10^{-11}$\,m$^2$ is the cross sectional area of the diode and 
$J_{\rm tot}$ is the total current density. In the simulations we use
$R = 5\,\Omega$, $L = 50\,$pH, and $C = 10\,$fF.

According to the Kirchhoff circuit laws, we have
\begin{equation}\label{eq:network_equations}
  U_e = U_R + U_{\rm RTD} + U_L, \quad U_{\rm RTD} = U_C, \quad
  I_L = I_R, \quad I_L = I_{\rm RTD} + I_C.
\end{equation}
Combining \eqref{eq:current_voltage_relationships} and \eqref{eq:network_equations}, 
we find that
\begin{align*}
  C\dot U_{\rm RTD} &= C\dot U_C = I_C = I_L - I_{\rm RTD}, \\
  L\dot I_L &= U_L = U_e-U_R-U_{\rm RTD} = U_e - RI_R - U_{\rm RTD}
  = U_e - RI_L - U_{\rm RTD}.
\end{align*}
Consequently, we obtain a system of two coupled ordinary differential equations,
\begin{equation}\label{eq:coupled_ode}
  \frac{d}{dt}\begin{pmatrix} U_{\mathrm{RTD}} \\ I_L \end{pmatrix}
  = \begin{pmatrix} 0 & \frac{1}{C} \\ -\frac{1}{L}	& -\frac{R}{L} \end{pmatrix}
  \begin{pmatrix} U_{\mathrm{RTD}} \\ I_L \end{pmatrix}
  + \begin{pmatrix} -\frac{1}{C} I_{\mathrm{RTD}} \\ \frac{1}{L} U_e(t)\end{pmatrix}.
\end{equation}

The time-step size $\triangle t$ is very small compared to the time scale
of the variation of the potential energy and the variation of the current flowing
through the diode. Hence, using the same time step for the time integration of
\eqref{eq:coupled_ode}, we can resort to an explicit time-stepping method.
We choose the simplest one, the explicit Euler method.
Alternatively, one may employ an implicit method, but we observed that both methods yield essentially the same results.

\begin{figure}[htb]
\begin{centering}
\includegraphics[width=150mm]{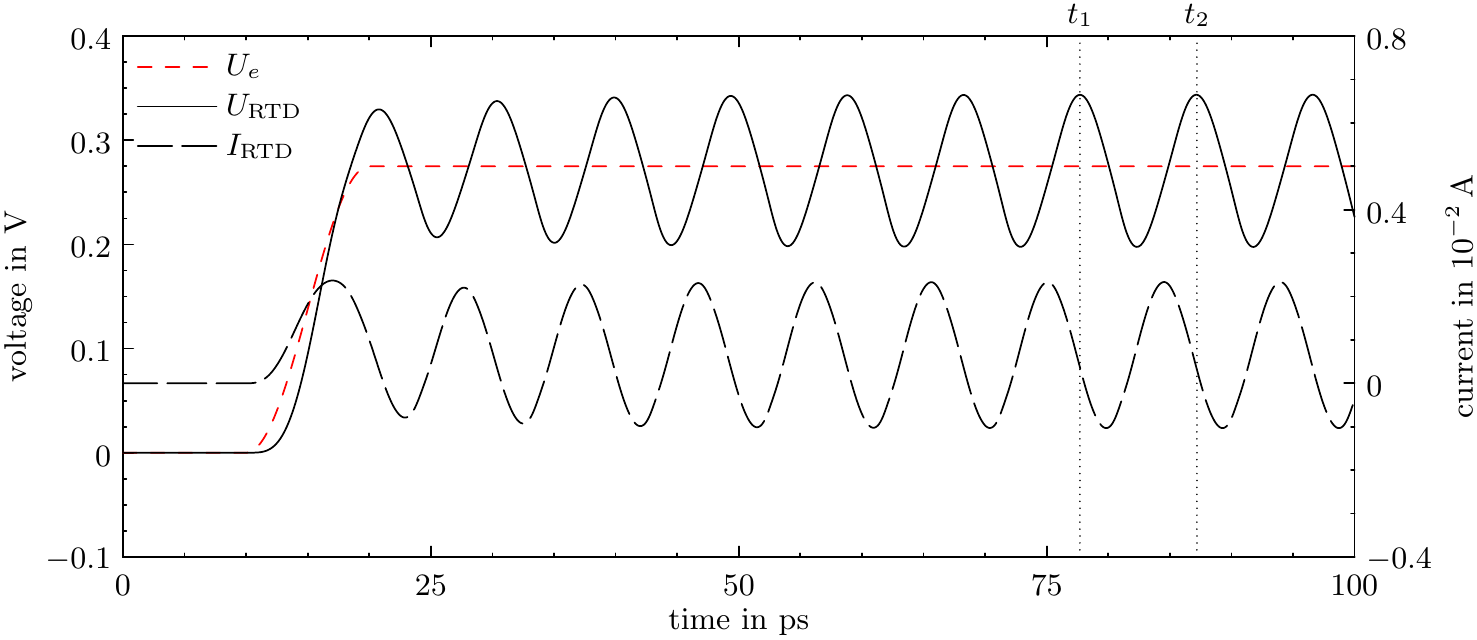}
\caption{
First circuit simulation: Voltage $U_{\rm RTD}$ and
current $I_{\rm RTD}$ through the resonant tunneling diode versus time.
}
\label{fig:oscillator_pde_U_RTD_and_I_RTD_of_time}
\end{centering}
\end{figure}

\begin{figure}
\begin{center}
\includegraphics[width=65mm]{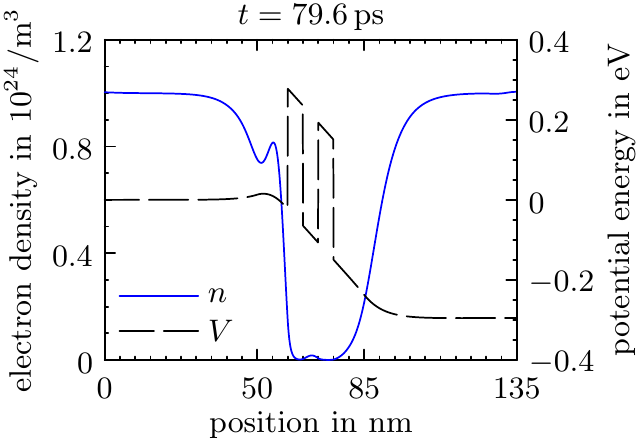}
\hspace{2.5mm}
\includegraphics[width=70mm]{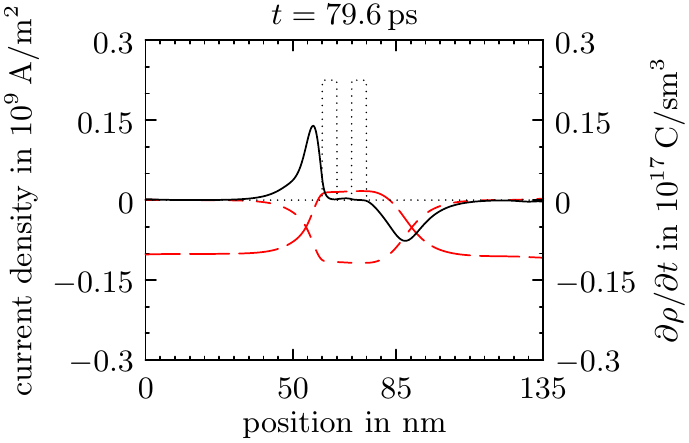}\label{fig:oscillation_details_part_1_a}
\\
\vspace{2.5mm}
\includegraphics[width=65mm]{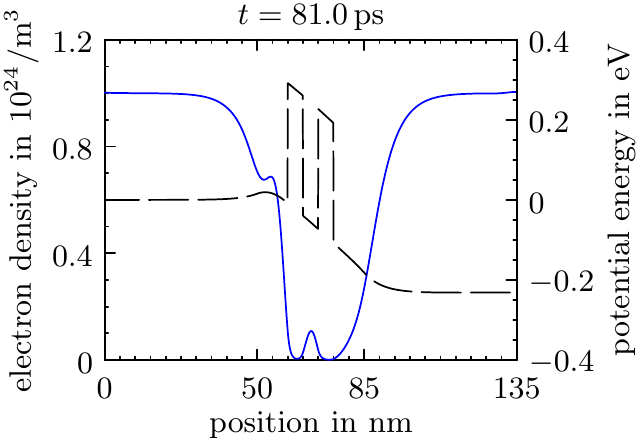}
\hspace{2.5mm}
\includegraphics[width=70mm]{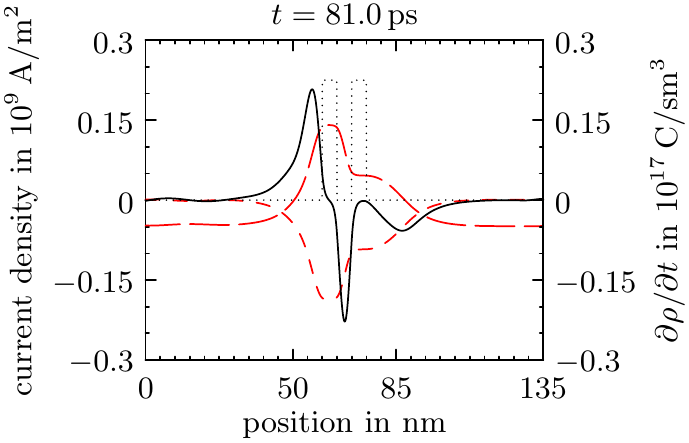}
\label{fig:oscillation_details_part_1_b}
\\
\vspace{2.5mm}
\includegraphics[width=65mm]{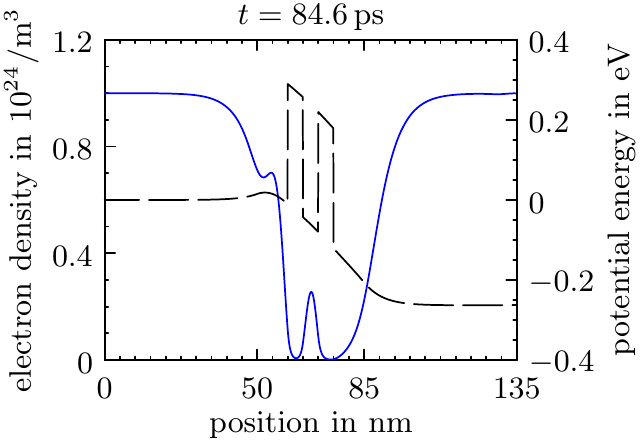}
\hspace{2.5mm}
\includegraphics[width=70mm]{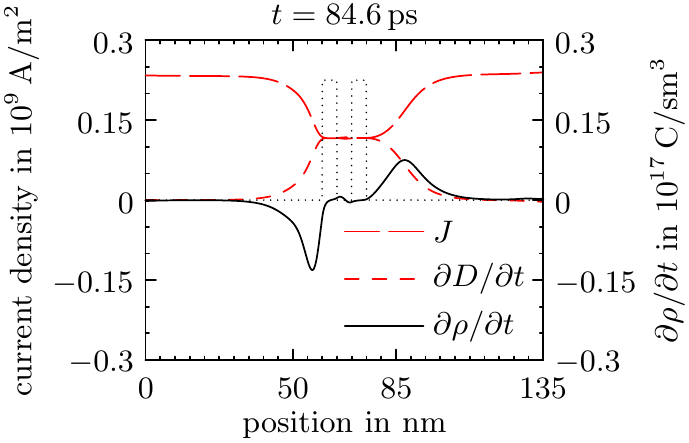}\label{fig:oscillation_details_part_1_c}
\\
\vspace{2.5mm}
\includegraphics[width=65mm]{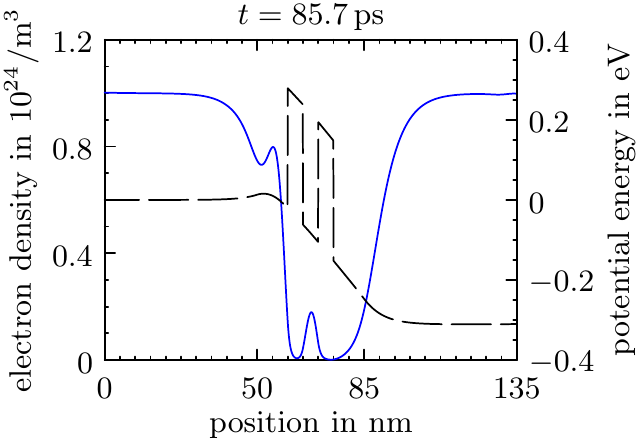}
\hspace{2.5mm}
\includegraphics[width=70mm]{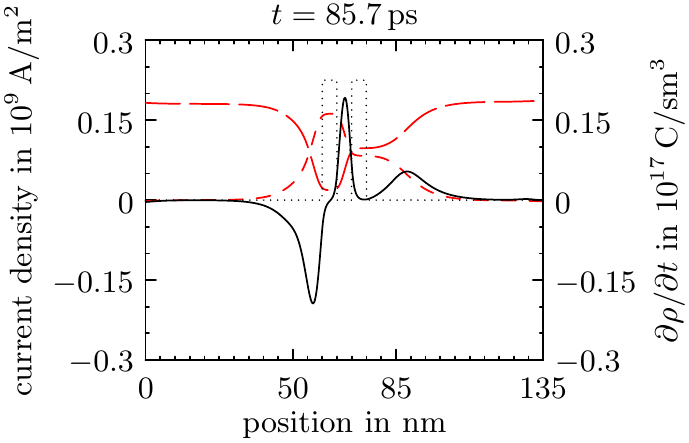}\label{fig:oscillation_details_part_1_d}
\end{center}
\caption{First circuit simulation: Electron density, potential energy, current densities, and 
variation of the electron density versus position in the RTD at four different
times.}
\label{fig:oscillation_details}
\end{figure}

\medskip\noindent
{\em First circuit simulation.}
In the first simulation, the RTD solver is initialized with the steady state
corresponding to $U_{\rm RTD}(t)=0$ for all $t\le 0$. The external voltage
$U_e$ is assumed to be zero for $t\le 0$, and the initial conditions
for \eqref{eq:coupled_ode} are $U_{\rm RTD}(0)=0$ and $I_L(0)=0$. 
For $t \in [10,20]$\,ps, the external voltage is increased smoothly to
0.275\,V and then kept constant
(see Figure \ref{fig:oscillator_pde_U_RTD_and_I_RTD_of_time}).
This value is between the voltages where the stationary current density reaches
its local maximum and minimum (see Figure \ref{fig.current_density_of_voltage_curves}).
The time evolution of the voltage and the current at the RTD are depicted
in Figure \ref{fig:oscillator_pde_U_RTD_and_I_RTD_of_time}.
It is clearly visible that the system starts to oscillate.
Furthermore, the potential energy, electron density, current densities, and the temporal
variation of the total charge $\pa\rho/\pa t$ are shown in Figure
\ref{fig:oscillation_details}
for four different times from the interval $[t_1=77.7,t_2=87.2]$\,ps, 
which covers exactly one oscillation.
Around 2\,ps after the beginning of the period, the electron density within the quantum well in $[65,70]\,$nm becomes minimal (first row).
After some time, we observe a build-up of negative charge
in the quantum well with $\pa\rho/\pa t<0$ (second row).
At about $t=84.6$\,ps the electron density reaches its maximum value (third row).
Subsequently, the electrons leave the quantum well again and $\pa\rho/\pa t>0$ in
$[65,70]\,$nm (fourth row).
The frequency of the oscillations is approximately 105\,GHz
which corresponds qualitatively to frequencies observed in standard double-barrier tunneling diodes \cite{BSPMMM91}.
The temporal evolution of the physical quantities in the circuit
is animated in the video available at
\verb!http://www.asc.tuwien.ac.at/~mennemann/projects.html!.

\begin{figure}[htb]
\begin{centering}
\includegraphics[width=150mm]{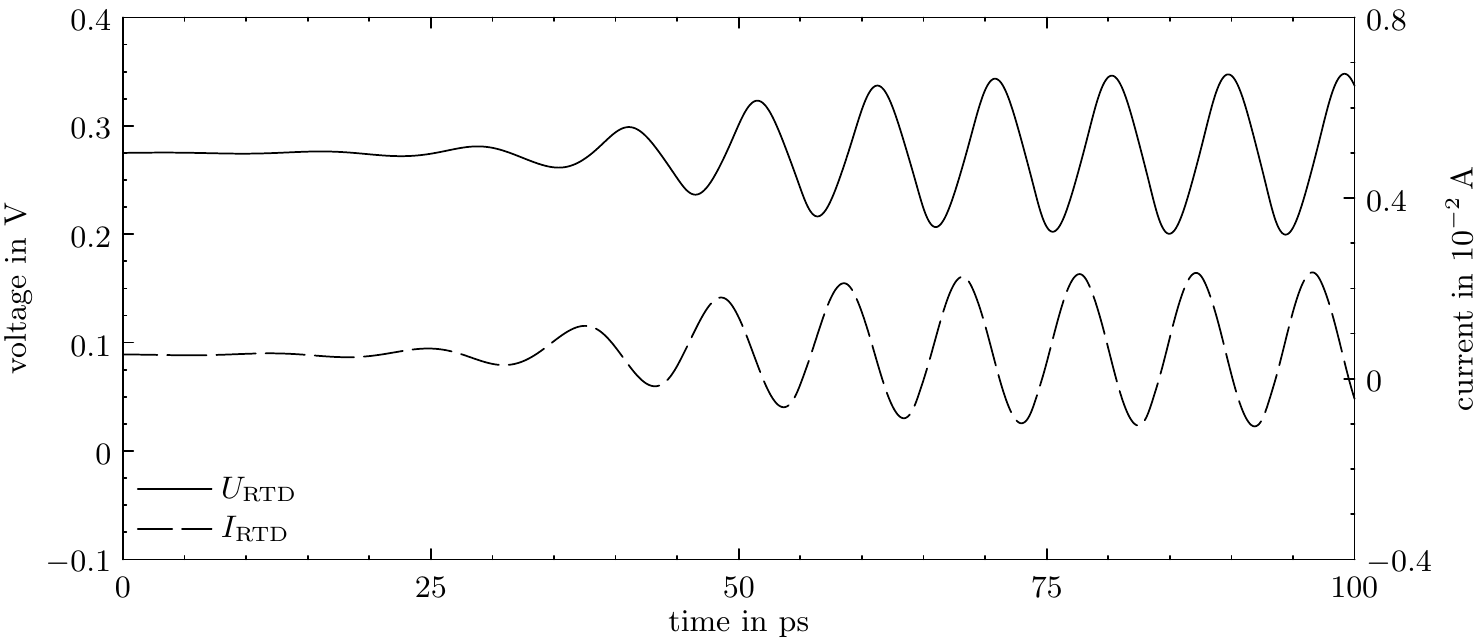}
\caption{Second circuit simulation: Applied voltage and current at the resonant tunneling diode
versus time.
}
\label{fig:oscillator_pde_instability_voltages_of_time}
\end{centering}
\end{figure}

\medskip\noindent
{\em Second circuit simulation.}
In this experiment, the external voltage $U_e$ is kept fixed for all times.
At times $t\le 0$, the circuit contains the voltage source, resistor, and RTD only.
We initialize the transient Schr\"odinger-Poisson solver with the steady state
corresponding to $U_{\rm RTD}(t)=0.275$\,V for all $t\le 0$. To compensate for
the voltage drop at the resistor, the external voltage is set to
$$
  U_e(t) = RI_{\rm RTD}(t) + U_{\rm RTD}(t), \quad t\le 0.
$$
At time $t=0$, the capacitor and inductor are added to the circuit. In order to avoid
discontinuities in the voltages, we charge the capacitor with the same voltage
wich is applied at the RTD before the switching takes place,
$U_C(t) = U_{\rm RTD}(t)$ for $t\le 0$.
For similar reasons, we set the current flowing through the inductor to the
current flowing through the RTD, $I_L(t)=I_{\rm RTD}(t)$ for $t\le 0$.
This configuration corresponds to the equilibrium state. Therefore,
one would expect that the system remains in its initial state for all time.
However, the equilibrium is unstable and a small perturbation will drive the
system out of equilibrium. In fact, numerical inaccuracies suffice to 
start the oscillator. However, we accelerate the transient phase by perturbing
$I_L(t)$ by the value $5\cdot 10^{-6}$\,A for $t\le 0$. 
The numerical result is presented in
Figure \ref{fig:oscillator_pde_instability_voltages_of_time}. 
The simulation took less than 4 hours computing time on an Intel Core 2
Quad Core Q9950 with $4\times 2.8$\, GHz. 

\section*{Acknowledgements}
The first two authors acknowledge partial support from   
the Austrian Science Fund (FWF), grants P20214, P22108, and I395, and    
the Austrian-French Project of the Austrian Exchange Service (\"OAD).


\end{document}